\documentclass[iop]{emulateapj}
\usepackage{amssymb,amsmath}
\usepackage{ulem}
\usepackage{color}
\usepackage[colorlinks=true,linkcolor=blue,anchorcolor=blue,citecolor=blue,urlcolor=blue]{hyperref}
\slugcomment{Last updated: \today}
\shorttitle{The Abundance  of  Intermediate-redshift Compact Galaxies}
\shortauthors{Damjanov et al.}

\begin{document}

\title{The Number Density of Quiescent Compact Galaxies at Intermediate Redshift}

\author{Ivana Damjanov\altaffilmark{1}, Ho Seong Hwang\altaffilmark{2}, Margaret J. Geller\altaffilmark{2}, and Igor Chilingarian\altaffilmark{2,3}}
\altaffiltext{1}{Harvard-Smithsonian Center for Astrophysics, 60 Garden Street, Cambridge, MA 02138; \email{idamjanov@cfa.harvard.edu}}
\altaffiltext{2}{Smithsonian Astrophysical Observatory, 60 Garden St., Cambridge, MA 02138}
\altaffiltext{3}{Sternberg Astronomical Institute, Moscow State University, 13 Universitetsky prospect, 119992 Moscow Russia}

\begin{abstract}

Massive compact systems at $0.2<z<0.6$ are the missing link between the predominantly compact population of massive quiescent galaxies at high redshift and their analogs and relics in the local volume. The evolution in number density of these extreme objects over cosmic time is the crucial constraining factor for the models of massive galaxy assembly. We select a large sample of $\sim200$ intermediate-redshift massive compacts from the BOSS spectroscopic dataset by identifying point-like SDSS photometric sources with spectroscopic signatures of evolved redshifted galaxies. A subset of our targets have publicly available high-resolution ground-based images that we use to augment the dynamical and stellar population properties of these systems by their structural parameters. We confirm that all BOSS compact candidates are as compact as their high-redshift massive counterparts and less than half the size of similarly massive systems at $z\sim0$. We use the completeness-corrected numbers of BOSS compacts to compute lower limits on their number densities in narrow redshift bins spanning the range of our sample. The abundance of extremely dense quiescent galaxies at $0.2<z<0.6$ is in excellent agreement with the number densities of these systems at high redshift. Our lower limits support the models of massive galaxy assembly through a series of minor mergers over the redshift range $0<z<2$.

\end{abstract}


\section{Introduction}\label{intro}

The most intriguing feature of the quiescent galaxy population at high redshift  is the high fraction of massive compact galaxies that are up to $\sim5$ times smaller than similarly massive passively evolving systems in the local Universe. In recent years a large suite of studies describe the extreme sizes and other structural properties of these quiescent objects at $z\gtrsim1$ \citep[e.g.,][]{Daddi2005, Longhetti2007,Trujillo2007, Toft2007, Zirm2007, Cimatti2008, vanDokkum2008, Buitrago2008, Damjanov2009, Damjanov2011, VanderWel2011, Bruce2012, Ryan2012, McLure2012, Chang2013}. 

In spite of the challenges of measuring sizes and stellar masses for these distant compact galaxies \citep[e.g.,][]{Mancini2010, Muzzin2009}, the compactness of high-redshift quiescent systems is a firmly established result \citep[e.g.,][]{ Szomoru2012, Cassata2010, VanderWel2012, Mosleh2013}. Dynamical mass measurements at high redshift \citep{Cappellari2009, Newman2010, Toft2012, vandeSande2013, Belli2014} for the brightest and most massive quiescent systems, confirm that $\gtrsim50\%$ of galaxies with $M_{dyn}>2\times10^{10}\, M_\sun$ are at most half the size of their massive local counterparts.  

The observed trend of galaxy size growth presents a challenge for models of massive galaxy evolution. In the presently favored scenario individual quiescent galaxies expand through a series of minor mergers (i.e., accretions of small gas-poor satellites) that build extended stellar envelopes around compact galaxy cores \citep[e.g,][]{Naab2009, vanDokkum2010, Oser2012}. In synergy with the continuous addition of larger newly quenched galaxies to the quiescent population \citep[progenitor bias; e.g,][]{vanDokkum2008, vanderWel2009, Saglia2010, Lopez-Sanjuan2012, Carollo2013, Cassata2013}, this scenario may explain the increasing average size of  quiescent galaxies with decreasing redshift. 

The stochastic nature of the merging processes requires that a non-negligible number of massive quiescent systems at each redshift between $z\sim2$ and $z=0$ be old and compact galaxies and that the fraction of such systems  decreases with time. Based on the Millennium $N-$body simulation, \citet{Quilis2013} show that if mass increases between 10\% and  30\% from $z \sim2$ to $z\sim0$, the fraction of massive galaxy relics at $z \sim 0$ should be only $0.1-1\%$. 

Searches for compact galaxies in the local volume yield a confusing picture. Studies based on the Sloan Digital Sky Survey (SDSS) show that the fraction of all compact systems among massive quiescent galaxies at $z<0.2$ is well below the model predictions,
particularly for old red sequence systems, the only possible candidates for high-redshift compact galaxy relics \citep{Trujillo2009, Taylor2010}. The low estimates of this population  dominated by an old stellar population challenges the proposed theory for massive galaxy assembly. These observations suggest that dry merging may not be the only mechanism driving galaxy size growth with redshift \citep[e.g.,][]{Nipoti2009, Hopkins2010}.   

Data from the WINGS survey of nearby clusters \citep{Fasano2006} paint a different picture of compact galaxies at $z \sim0$.  Their number density estimate is  two orders of magnitude above the estimates based on SDSS dataset. The compact galaxy candidates in this study reside exclusively in cluster environment \citep{Valentinuzzi2010}.  Yet another spectroscopic galaxy sample representing the general field \citep[PM2GC survey,][]{Calvi2011} contains a very high number density of compact systems, similar to their abundances at high redshift \citep{Poggianti2013a}. These extreme number densities lie at or above the upper limits of model predictions.
   
The population of dense passively evolving galaxies at intermediate redshift ($0.2<z<0.8$) is a crucial  link between the compact systems that dominate the massive quiescent galaxy population at high redshift and their analogs and relics in the local volume. The analysis of structural and dynamical properties of quiescent galaxies at $0.2<z<0.9$, based predominantly on cluster data \citep[EDisCS,][]{White2005}, shows that progenitor bias can significantly decrease the amount of size evolution for massive ellipticals in this redshift range \citep{Saglia2010}. A photometric study of massive quiescent galaxies within the COSMOS field shows that the number density of the most compact systems does not evolve dramatically from $z\sim1$ to $z\sim0.2$ \citep{Carollo2013}. Although mostly focused on $z \gtrsim 0.6$, analysis of the GOODS fields gives similar results \citep{Cassata2011, Cassata2013}.  However, a small spectroscopic sample of massive compact systems at $z\sim0.5$ shows that only a fraction could have formed at $z>2$ \citep{Stockton2014,Hsu2014}. Clearly, larger samples of compact galaxies with known spectroscopic properties and high-quality imaging are necessary for connecting their morphology, dynamical properties, and assembly histories. A comprehensive view of the evolution in the abundance of these systems requires number densities based on spectro-photometric samples spanning the intermediate redshift range. 
  
In \citet{Damjanov2013} we selected a large sample of $\sim700$ intermediate-redshift compact candidates from the spectro-photometric SDSS database by combining point-like morphologies with the spectroscopic signatures of redshifted passively evolving galaxies. We confirm extreme compactness for a small subset of 9 compact candidates serendipitously observed by the HST. These publicly-available datasets provide us with all the information we need to identify these very unusual, rare objects and to measure their structural, dynamical, and stellar population parameters.  The heterogeneous parent sample precludes an estimate of the number density of these compact objects. However, even this small sample of intermediate redshift compact SDSS galaxies demonstrates that these systems inhabit a range of environments: out of nine targets, we found only one rich cluster member \citep{Damjanov2013}.

Here we describe a new sample of $\sim200$ compact galaxies drawn from the Baryon Oscillation Spectroscopic Survey \citep[BOSS,][]{Eisenstein2011} and cross-matched with publicly available high-quality Canada-France-Hawaii Telescope (CFHT) MegaCam imaging. We use this unique set of compact galaxy 
candidates to assess, for the first time, number densities of these systems in narrow redshift bins covering 3.3 Gyrs (25\%) of the cosmic history between $z=0.2$ and $z=0.6$. In Section~\ref{select} we describe the selection process and present galaxy sample. We display structural properties for a subsample of compact candidates with available high-quality imaging in Section~\ref{ratify}. We combine these structural properties with velocity dispersions we measure from  BOSS spectra and compare our targets with massive quiescent systems at $z=0$ and $z>1$ in Section~\ref{fundamental}. We show that this number of confirmed compact intermediate-redshift systems implies that most of our BOSS candidates are indeed compact. Thus we can use these targets to compute lower limits on compact galaxy number densities  in the intermediate-redshift range  and to augment  the observational picture of the number density evolution of these extreme systems  with cosmic time (Sections~\ref{spec} and~\ref{res}). We compile all of the number density estimates and compare them with models in Sections~\ref{comp} and~\ref{model}. We adopt a cosmological model with $\Omega_{\Lambda}=0.7$, $\Omega_{M} = 0.3$, and $H_0 = 70$~km~s$^{-1}$~Mpc$^{-1}$, and quote magnitudes in the AB system throughout. Error bars correspond to $1\sigma$ errors throughout.
    
\section{Compact Galaxy Candidates in BOSS Dataset}\label{select}

We use the BOSS DR10 dataset  \citep{Ahn2014} and employ the selection criteria of \citet{Damjanov2013}  to select compact galaxy candidates in the redshift range $0.2<z<0.6$. We search for stellar-like objects with spectroscopic redshifts in the redshift range of interest (22,288 sources) and exclude all spectroscopically confirmed quasars (14,108 sources) and actively star-forming galaxies (by requiring the equivalent width of the $[$OII$]\lambda\lambda3726,3729$ emission line to be EW$[$O II$]<5$~\AA; 7,299 sources excluded). In addition, we inspect every pre-selected spectrum to identify a 4000~\AA~break and several absorption features including Balmer series, Ca H+K and G-band. We also apply a cut in the inferred dynamical mass at $M_{dyn}=10^{10}\,M_\sun$ to construct a sample of compact quiescent systems as massive as high- and low-redshift comparison samples (see Sections~\ref{fundamental} and~\ref{comp}). The final list of BOSS compact galaxy candidates at redshift $0.2<z<0.6$ contains 198 objects. BOSS initially selected $95\%$ of our sample for spectroscopic followup as quasar (rather than galaxy) candidates.

We reanalyze spectra of our BOSS targets to measure redshift, velocity dispersion, age, and metallicity following the procedure of \citet{Damjanov2013}. The core of our fitting routine is the comparison of each galaxy spectrum with a grid of {\sc pegase.hr} \citep{LeBorgne+04} simple stellar population (SSP) models based on the MILES stellar library \citep{SanchezBlazquez+06} using the {\sc nbursts} pixel space fitting technique \citep{CSAP07,CPSK07}.  Firstly we convolve the SSP model grid covering a wide range of ages and metallicities with the instrumental response of the BOSS spectrograph. We then convolve these SSP models again with a Gaussian line-of-sight velocity distribution and multiply by a smooth low-order continuum polynomial in order to absorb flux calibration errors in both models and the data. We choose the best-fitting SSP (or a linear combination of two) by interpolating a grid in age and metallicity. 

Here we present single SSP fitting of the BOSS spectra: we choose the best-fitting stellar population template from the grid of models in age-metallicity space explored during the nonlinear minimization procedure. The errors in the resulting stellar population parameters and internal kinematical properties are formal statistical errors based on the Levenberg-Marquardt least-squares algorithm.  \citet{Chilingarian2008}~and~\citet{Chilingarian2009} discuss the strengths of this approach in more details and \citet{Fabricant2013} demonstrate that its application to BOSS spectra produces stable stellar kinematic measurements without any significant systematic uncertainties. The errors in stellar age and metallicity do not include uncertainties introduced by our choice of stellar population models or star formation history representation. For example, high signal-to-noise spectra can be successfully fitted using a combination of several SSP templates. Alternatively, a grid of models for exponentially declining star formation histories can be used instead of SSPs. A more detailed study of stellar populations in quiescent galaxies is beyond the scope of this paper.

\begin{figure*}
\begin{centering}
\includegraphics[scale=0.35]{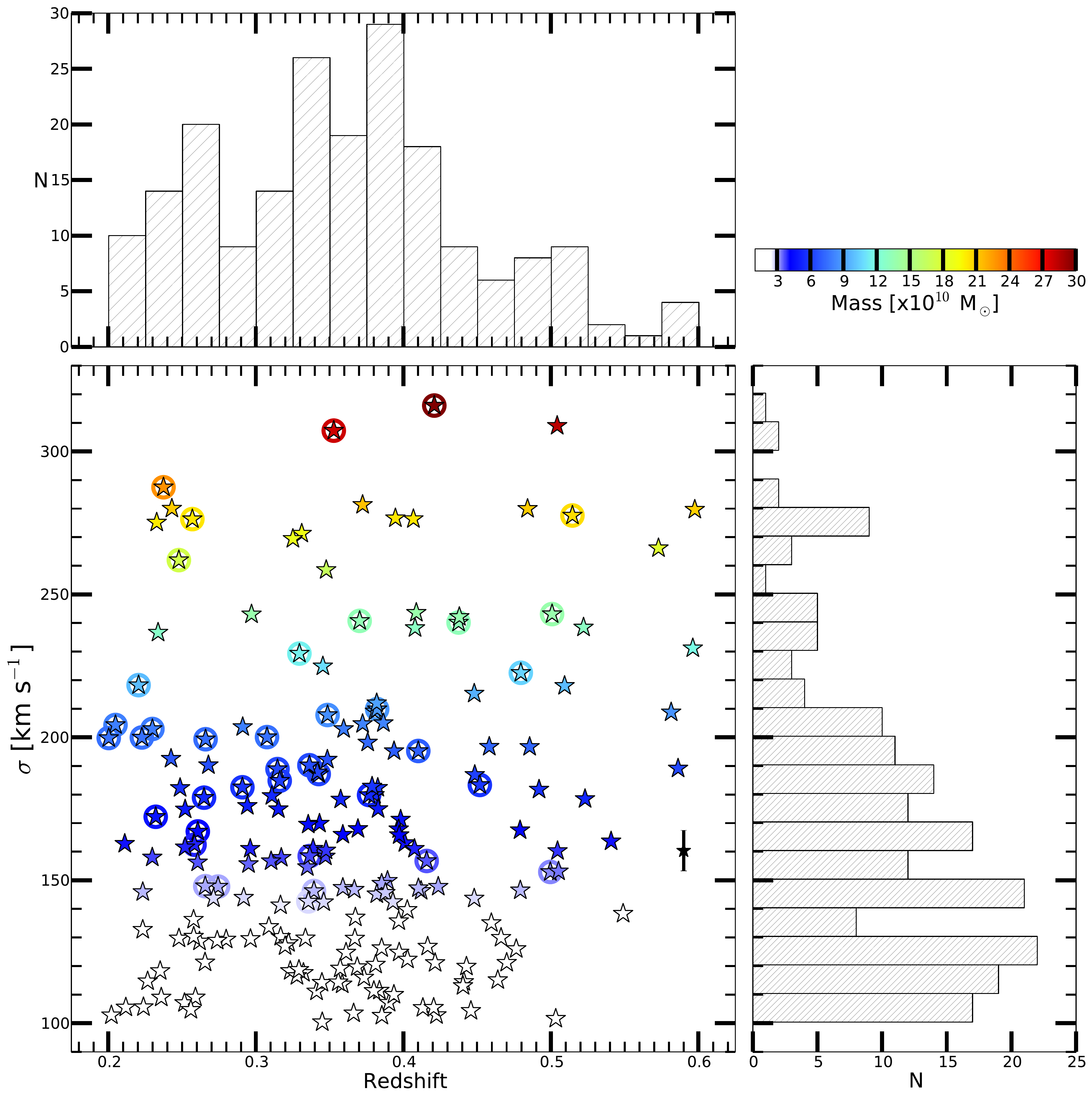}
\caption{ Velocity dispersion and redshift distribution of the intermediate-redshift compact galaxy candidates selected mainly from the BOSS quasar survey. Stars in the main panel are color-coded by galaxy dynamical mass, derived from measured velocity dispersion (Section~\ref{fundamental}, Eq.~\ref{eq: sigma}). Circled symbols denote compact candidates formed at 
$z_\mathrm{form}>2$. Black symbol with errorbar shows the average fractional error of the velocity dispersion measurement placed at the median velocity dispersion for our sample. Central panel displays the incompleteness of our magnitude limited sample for low-mass objects at $z\gtrsim0.5$. Augmented by the fact that BOSS quasar selection function excludes majority of red objects (at high masses and/or high redshifts), this plot confirms that the number densities of intermediate-redshift compact galaxies we present in Section~\ref{nd} are hard lower limits on the abundance of these extreme systems in this redshift range. \label{f1}}
\end{centering}
\end{figure*}

Table~\ref{tab1} lists all of the spectroscopic properties for our targets. Figure~\ref{f1} shows distributions of two parameters we use in our analysis, the velocity dispersion ($\sigma$) and the redshift ($z$). Median values of these two parameters for the BOSS sample of quiescent compact galaxy candidates are $\widetilde{\sigma}=160$~km s$^{-1}$ and $ \widetilde{z}=0.36$. Points representing our targets in the velocity dispersion -- redshift parameter space (central panel of Figure~\ref{f1}) are color coded based on their dynamical mass estimated from the velocity dispersion -- dynamical mass relation for a subsample of object with measured sizes (Section~\ref{fundamental}). Dynamical masses of our compact galaxy sample span a range of values: $1\times10^{10}$~M$_{\sun} \leqslant M_{dyn} \leqslant 3\times10^{11}$~M$_\sun$, with a median of $\tilde{M}_{dyn}=4\times10^{10}$~M$_\sun$ and with 22\% of the sample objects having ${M}_{dyn}\geqslant8\times10^{10}$~M$_\sun$.

\begin{figure*}
\begin{centering}
\includegraphics[scale=0.6]{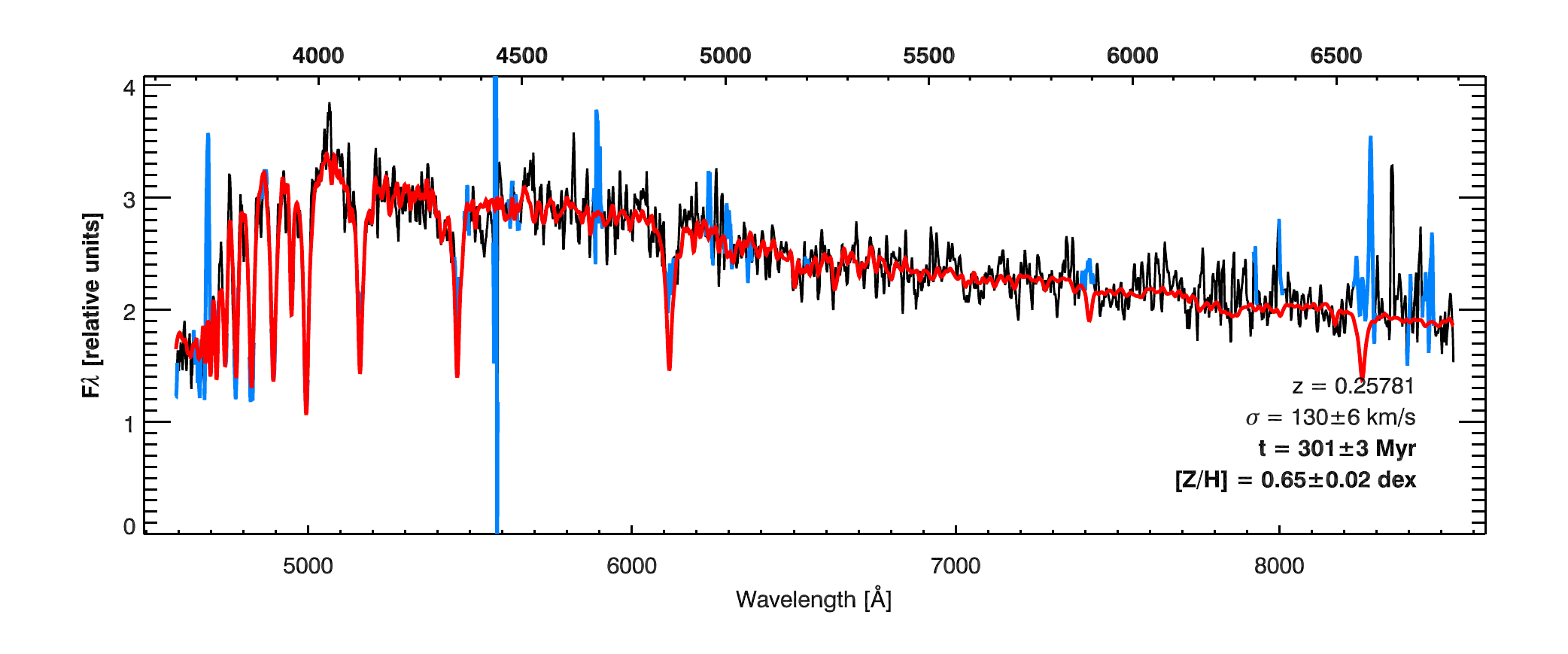}
\includegraphics[scale=0.6]{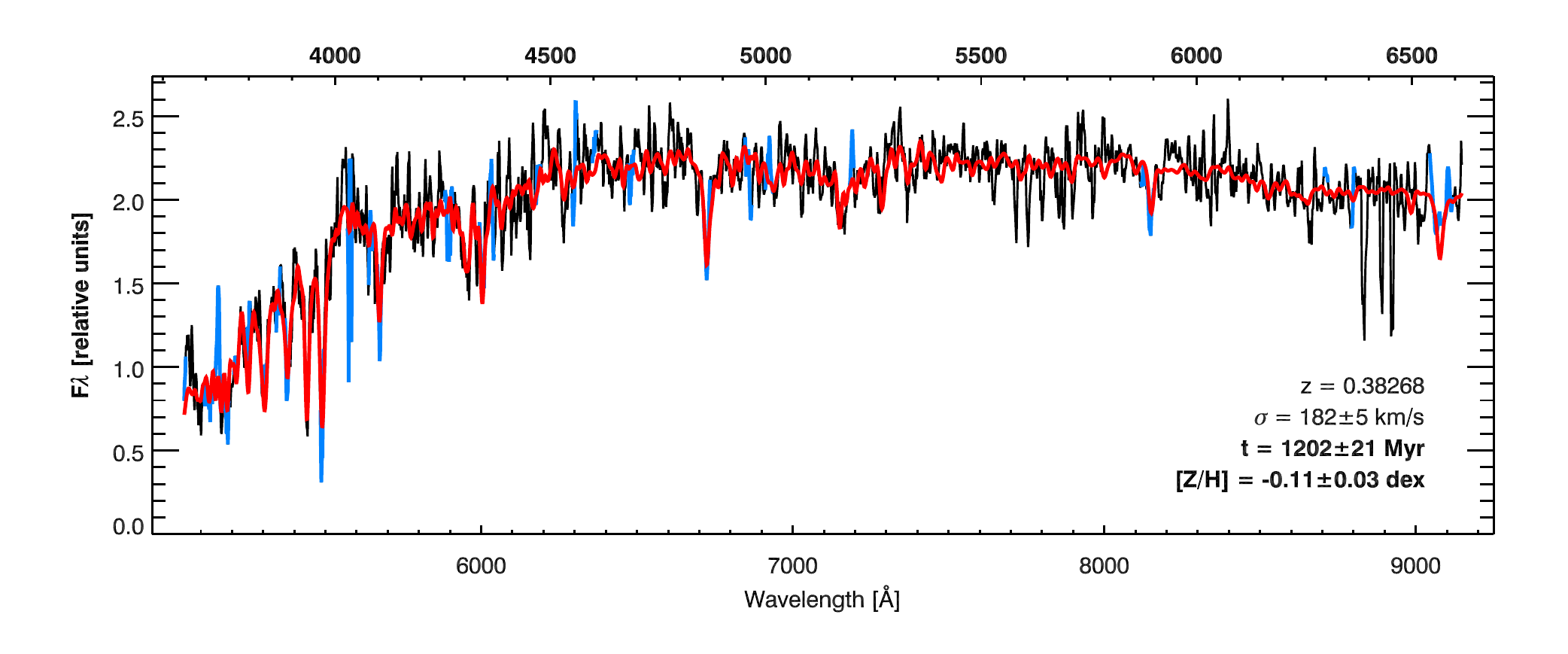}
\includegraphics[scale=0.6]{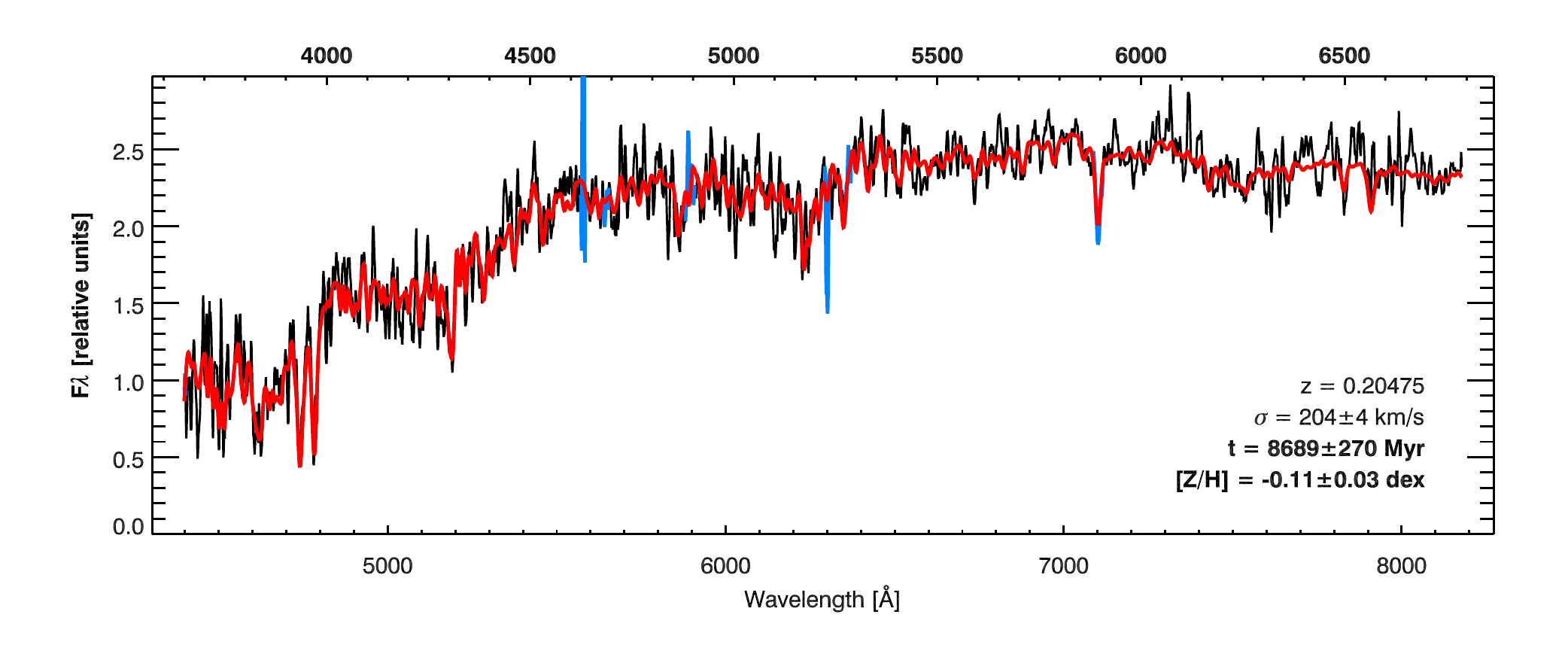}
\caption{Three examples showing typical spectra of: young (top panel), intermediate (E+A, central panel), and old (bottom panel) compact intermediate-redshift galaxies we select from the parent sample of BOSS quasar targets. Each panel show the BOSS spectrum (black), the best-fit SSP spectrum (red), and regions excluded form the fit (blue). Labels display redshift, velocity dispersion, age, and metallicity for each galaxy example. The bottom horizontal axis displays observed wavelength range we use for the SSP fitting, and the upper horizontal axis shows wavelength in the galaxy rest-frame. \label{f2}}
\end{centering}
\end{figure*}

The age of the best fitting SSP model is an important spectroscopic parameter for characterizing our sample. Because of the `age-metallicity degeneracy'  age is a poorly constrained fitting parameter. Thus there may be a confusion between old, metal-poor galaxies and young, metal-rich ones \citep{Worthey1994}. Figure~\ref{f2} displays three example spectra from our dataset, along with the best-fit SSP models and their parameters. As does velocity dispersion (and subsequently dynamical mass), the ages of our compact galaxy candidates cover a wide range. Fewer than 10\% of our targets have ages comparable with `extreme post-starburst systems' described in Damjanov et al. 2013, and almost 50\% span the age range of E+A galaxies \citep[500 Myr -- 2 Gyr; e.g.,][]{Poggianti1999}. 

Ages of $\sim20\%$ of our targets place their formation redshift at $z_\mathrm{form}>2$ (circles in Figure~\ref{f1}).  With dynamical masses $1.2\times10^{10}$~M$_{\sun} \lesssim M_{dyn}\lesssim 3\times10^{11}$~M$_\sun$ (i.e, in the range of the high-redshift samples we use for comparison in Sections~\ref{fundamental} and~\ref{comp}), the subsample of old compact galaxies at $0.2<z<0.6$ represents a set of candidates for relics of the `red nuggets' found at $z>1$.                

\begin{deluxetable*}{ccccccc}
\tabletypesize{\scriptsize}
\tablecaption{Spectroscopic properties of compact galaxy candidates selected from BOSS \label{tab1}}
\tablewidth{0pt}
\tablehead{\colhead{SDSS objID} & \colhead{$z$} & \colhead{$\alpha$} & \colhead{$\delta$} & \colhead{$\sigma$} & \colhead{Age} & \colhead{[Z/H]}\\
\colhead{} & \colhead{} & \colhead{[$\arcdeg$]} & \colhead{[$\arcdeg$]} & \colhead{[km s$^{-1}$]} & \colhead{[Myr]} & \colhead{[dex]} \\
\colhead{(1)} & \colhead{(2)} & \colhead{(3)} & \colhead{(4)} & \colhead{(5)} &\colhead{(6)} & \colhead{(7)}\\ 
}
\startdata
1237663784202207380 & 0.224 & 11.0073 & 0.114363 & 106 $\pm$ 4 & 4123 $\pm$ 225 & -0.26 $\pm$ 0.05 \\
1237679340567527799 & 0.413 & 31.9741 & -6.30756 & 105 $\pm$ 12 & 1562 $\pm$ 39 & -1.11 $\pm$ 0.04 \\
1237663784216167033 & 0.353 & 42.9603 & 0.150814 & 307 $\pm$ 7 & 9029 $\pm$ 413 & -0.05 $\pm$ 0.04 \\
1237660241384964734 & 0.236 & 44.4049 & 0.475132 & 109 $\pm$ 4 & 1614 $\pm$ 41 & -0.49 $\pm$ 0.05 \\
1237655126076096784 & 0.388 & 168.6826 & 5.281579 & 146 $\pm$ 6 & 748 $\pm$ 14 & -0.19 $\pm$ 0.05 \\
1237654605329859282 & 0.505 & 169.5945 & 4.815115 & 153 $\pm$ 11 & 642 $\pm$ 18 & -0.24 $\pm$ 0.08 \\
1237662238540431745 & 0.369 & 185.5511 & 10.837048 & 120 $\pm$ 2 & 1952 $\pm$ 19 & -0.1 $\pm$ 0.01 \\
1237661972797719126 & 0.367 & 188.5982 & 8.325914 & 137 $\pm$ 5 & 1035 $\pm$ 10 & 0.62 $\pm$ 0.01 \\
1237662262167339212 & 0.258 & 210.3415 & 7.851378 & 162 $\pm$ 4 & $<10450$ & -0.24 $\pm$ 0.04 \\
1237651752414675608 & 0.448 & 214.2453 & 1.019469 & 242 $\pm$ 13 & 1434 $\pm$ 42 & -0.54 $\pm$ 0.05 \\
1237663543685414961 & 0.408 & 335.8087 & 0.405848 & 238 $\pm$ 6 & 2570 $\pm$ 61 & 0.27 $\pm$ 0.04 \\
1237663543685545989 & 0.409 & 336.0761 & 0.358809 & 244 $\pm$ 3 & 3387 $\pm$ 68 & 0.12 $\pm$ 0.02 \\
1237660025032868014 & 0.358 & 342.2582 & -0.697181 & 178 $\pm$ 5 & 4334 $\pm$ 184 & -0.27 $\pm$ 0.04 \\
1237650804268400981 & 0.345 & 153.465 & -2.351702 & 100 $\pm$ 6 & 1074 $\pm$ 27 & -0.18 $\pm$ 0.06 \\
...&...&...&...&...&...&...
\enddata
\tablecomments{Columns: (1) SDSS object identification; (2) Redshift; (3) Right ascension; (4) Declination (5) Velocity dispersion; (6) Age of the best-fit SSP model; (7) Metallicity of the best-fit SSP model.}
\tablenotetext{}{The internal kinematics and stellar population errors are formally computed statistical errors and do not include possible systematic uncertainties} 
\tablenotetext{}{A portion of Table~\ref{tab1}, presenting the spectroscopic properties of objects presented also in Table~\ref{tab2}, is shown here for guidance regarding its form and content.  The table is published in its entirety in the electronic edition of The Astrophysical Journal.}
\end{deluxetable*}

\section{Ratifying the compact nature of  unresolved galaxies in BOSS}\label{ratify}

The quality of SDSS III imaging, with a median Point Spread Function (PSF) width of $1\farcs43$ \footnote{\url{https://www.sdss3.org/dr9/imaging/other\_info.php}}, gives an upper limit  of $\sim2.3$ to $\sim4.6$~kpc in the redshift range $0.2<z<0.6$ on the effective radius of stellar-like objects spectroscopically classified as galaxies. To measure actual sizes of compact galaxy candidates in our sample we used the Astronomical Data Query Language (ADQL) to search the Canadian Astronomy Data Centre (CADC) collections for higher quality images  taken by the Canada-France-Hawaii Telescope (CFHT) imaging instrument MegaCam operating in the visible wavelength regime. \footnote{\url{http://www2.cadc-ccda.hia-iha.nrc-cnrc.gc.ca/en/search/?collection=CFHTMEGAPIPE&noexec=true}} 

Fourteen out of 198 compact candidates in our sample were previously observed using CFHT MegaCam under good seeing conditions (i.e.,  with Full Width at Half Maximum (FWHM) of the PSF $<0\farcs9$). We selected only combined images with total exposure times of $t_{exp}\gtrsim1800$~s corresponding to $5\sigma$ point source magnitude detection limit \footnote{\url{http://www2.cadc-ccda.hia-iha.nrc-cnrc.gc.ca/megapipe/docs/photometry.html\#limit}} of $26.8\leqslant m_{lim}(g')\leqslant27.3$ (5 objects), $25.9\leqslant m_{lim}(r')\leqslant26.2$ (2 objects), $25.6\leqslant  m_{lim}(i')\leqslant26$ (9 objects), and $24.9\leqslant m_{lim}(z')\leqslant25.4$ (2 objects). Three of our targets have CFHT MegaCam imaging in more than one filter, and we explore their color profiles in detail in Section~\ref{color}.  

\subsection{Surface brightness profile modeling}\label{galfit}

\begin{figure*}
\begin{centering}
\includegraphics[scale=0.45]{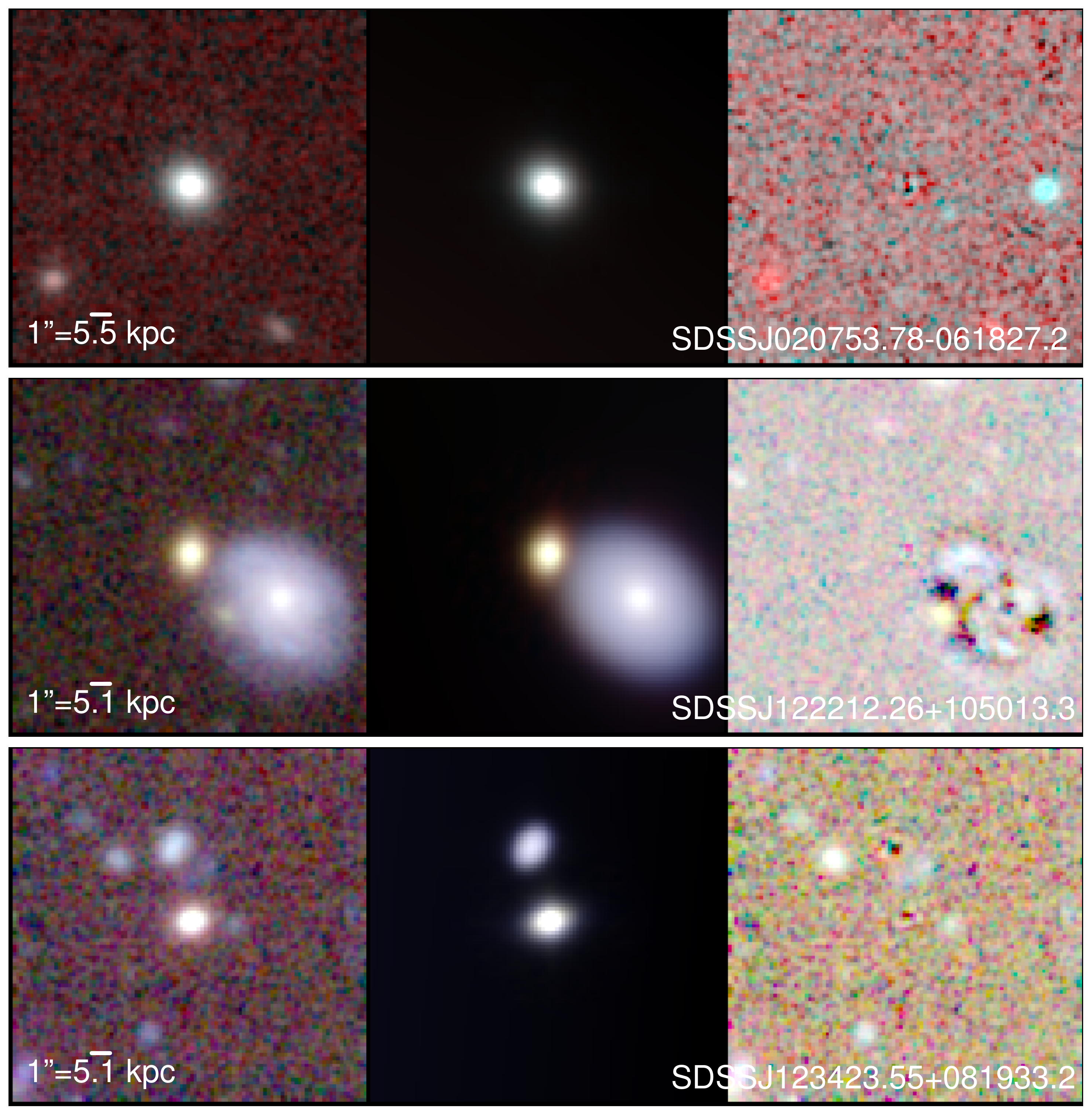}

\caption{Montage of the CFTH MegaCam images ({\it left}), best-fit GALFIT models ({\it center}), and residual maps ({\it right}) for 11 BOSS intermediate-redshift compacts that could be resolved with the FWHM(PSF)$\sim0\farcs5-0\farcs85$. Three galaxies with available multi-band images are presented with false color: SDSSJ0207-0618 (red - $z$ filter; blue - $i$ filter), SDSSJ1222+1050, and SDSSJ1234+0819 (red - $z$ filter; green - $i$ filter; blue - $g$ filter). \\ (An extended version of this figure is available in the online journal.)\label{f3}}
\end{centering}
\end{figure*}

We employ the GALFIT software v3.0.4 \citep{Peng2010} to fit the surface brightness profiles $\Sigma(r)$ of our BOSS/CFHT targets with 2D models described by parameters of a single \citet{SersicJoseLuis1968} profile:

\begin{equation} 
\Sigma(r)=\Sigma_e\times \mathrm{exp} \left[-\kappa\left(\left(\frac{r}{R_e}\right)^{\frac{1}{n}}-1\right)\right], \label{eq:sersic}
\end{equation}
 
\noindent where $R_e$ is the effective (or half-light) radius encompassing half of the total flux, $\Sigma_e$ is the surface brightness at $R_e$, $n$ is the S\'{e}rsic index describing the central concentration, and $\kappa$ is a normalization factor depending on $n$.  

Because the images retrieved from the CFHT database cover only a fraction ($\sim12\arcmin\times12\arcmin$) of the MegaCam field of view ($\sim1$~sq. degree) and the PSF can vary substantially, we do not use the information on average PSF size from the CFHT image headers. Instead, for each of the BOSS/CFHT compact galaxy candidates  we use a set of routines in the IRAF {\it daophot} package to identify and combine a number ($\sim10$) of bright unsaturated point sources in the vicinity of our target. This empirical PSF is one of the GALFIT input parameters. GALFIT convolves the empirical PSF with a single S\'{e}rsic models and compares the result with the observed surface brightness profile. Using the Levenberg-Marquardt algorithm the fitting procedure then tunes the parameters of the analytic model until the minimum sum of weighted pixel-scale deviations between the galaxy image and its model is reached. 

Figure~\ref{f3} shows the resulting best-fit models and Table~\ref{tab1} lists the parameters. Although CFHT MegaCam imaging is available for 14 compact galaxy candidates from BOSS, three targets are unresolved under the FWHM (PSF)$\lesssim0\farcs8$ seeing conditions. Figure~\ref{f3} displays a montage of galaxy surface brightness profiles, GALFIT best-fit models, and residual maps for the 11 resolved objects.  Deep high quality imaging confirms that all 14 targets are very compact, with circularized effective radii $R_{e,c}\lesssim2$~kpc. 

Three resolved galaxies have available CFHT MegaCam imaging in more than one filter. These targets, along with their best-fit models and residual maps in all available bands, are shown as false-color images in the montage. For each of these galaxies surface brightness profiles in all bands exhibit similar effective radii (Figure~\ref{f3}). In Section~\ref{color} we discuss radial profiles of these objects.  The similarity between sizes of compact BOSS galaxies measured in multiple filters is in agreement with the results of multi-band imaging of quiescent compact galaxies at $0.5<z<2.5$ \citep{Cassata2010, Szomoru2013}.

GALFIT reports the uncertainties that correspond to random errors only. To estimate the errors empirically, we construct a grid of GALFIT input files that include: a range of initial parameters for the fitting S\'{e}rsic profile, a range of sizes for the fitting region, neighboring objects that are either masked or fitted simultaneously, sky properties either determined from aperture photometry or set as input parameters for the fitting routine. As the authors of GALFIT note\footnote{\url{http://users.obs.carnegiescience.edu/peng/work/galfit/TFAQ.html\#errors}}, distributions of resulting best-fit parameters give more realistic error estimates than the ones provided by the fitting routine itself. If the intrinsic half-light radius (along major axis, $R_e$) of our target is smaller than the corresponding PSF, we assign the difference between the Half Width at Half Maximum (HWHM) of the PSF and the best-fit $R_e$ as the measurement error. We note that this conservative approach produces amplified errors, because the simulations show that the sizes of objects smaller than the image PSF tend to be overestimated \citep[e.g.,][]{Carollo2013}. For three unresolved compact galaxies, Table~\ref{tab2} lists the HWHM (PSF), in physical units at the redshift of each unresolved target, as upper limits on their sizes. 

In summary, we combed through the CFHT MegaCam database and found high quality images for $\sim7\%$ of the compact galaxy candidates selected from BOSS. We use GALFIT to obtain structural properties for these 14 objects and to confirm that they are all indeed very compact (with median circularized half-light radius $\tilde{R_{e,c}}\sim0.8$~kpc), with S\'{e}rsic indices typical for bulge-dominated systems ($n>2$).  Intermediate redshift compact galaxies, selected in the same manner from the SDSS photo-spectroscopic database and with available high-resolution HST images, exhibit very similar structural properties \citep[see Table 2 of][]{Damjanov2013}. In Section~\ref{fundamental} we compare structural and dynamical properties of these compact BOSS galaxies at intermediate redshift with similarly massive quiescent systems at $z\sim0$ and $z>1$.    

\begin{deluxetable*}{lccccccc}
\tabletypesize{\scriptsize}
\tablecaption{Structural properties of  BOSS compact galaxies at intermediate redshift with existing  CFHT imaging \label{tab2}}
\tablewidth{0pt}
\tablehead{
\colhead{SDSS objID} & \colhead{Name} & \colhead{$z$} & \colhead{$R_{e,c}$} & \colhead{$n$} & \colhead{$b/a$} & \colhead{Filter} & \colhead{HWFM(PSF)} \\
\colhead{} & \colhead{} & \colhead{} & \colhead{[kpc]} & \colhead{} & \colhead{} & \colhead{} & \colhead{[kpc]} \\
\colhead{(1)} & \colhead{(2)} & \colhead{(3)} & \colhead{(4)} & \colhead{(5)} & \colhead{(6)} & \colhead{(7)} &\colhead{(8)}\\
}
\startdata
1237663784202207380 & SDSSJ004401.76+000651.7 & 0.224 & 1.4 $\pm$ 0.14 & 5.66 $\pm$ 2.0 & 0.78 $\pm$ 0.04 & $i'$ & 1.02 \\
1237679340567527799 & SDSSJ020753.78-061827.2 & 0.413 & 0.75 $\pm$ 0.47 & 6.6 $\pm$ 0.25 & 0.75 $\pm$ 0.02 & $z'$ & 1.42 \\
1237663784216167033 & SDSSJ025150.47+000902.8 & 0.353 & 1.27 $\pm$ 0.62 & 2.59 $\pm$ 0.7 & 0.7 $\pm$ 0.02 & $i'$ & 1.34 \\
1237660241384964734 & SDSSJ025737.18+002830.4 & 0.236 & $<1.18$ & ... & ... & $i'$  & 1.18 \\
1237655126076096784 & SDSSJ111443.82+051653.6 & 0.388 & 0.66 $\pm$ 0.63 & 3.59 $\pm$ 0.47 & 0.34 $\pm$ 0.04 & $i'$ & 2.22 \\
1237654605329859282 & SDSSJ111822.67+044854.4 & 0.505 & $<2.78$ & ... & ... & $g'$  & 2.78 \\
1237662238540431745 & SDSSJ122212.26+105013.3 & 0.369 & 1.1 $\pm$ 0.11 & 2.07 $\pm$ 0.5 & 0.56 $\pm$ 0.02 & $i'$  & 1.39 \\
1237661972797719126 & SDSSJ123423.55+081933.2 & 0.367 & 0.42 $\pm$ 0.41 & 3.09 $\pm$ 0.4 & 0.36 $\pm$ 0.04 & $i'$  & 1.37 \\
1237662262167339212 & SDSSJ140121.96+075104.9 & 0.258 & 1.76 $\pm$ 0.25 & 3.77 $\pm$ 0.72 & 0.7 $\pm$ 0.09 & $r'$ & 1.25 \\
1237651752414675608 & SDSSJ141658.85+010110.1 & 0.438 & $<2.21$ & ... & ... & $r'$  & 2.21 \\
1237663543685414961 & SDSSJ222314.09+002421.0 & 0.408 & 2.02 $\pm$ 0.72 & 5.5 $\pm$ 2.1 & 0.88 $\pm$ 0.1 & $i'$  & 1.36 \\
1237663543685545989 & SDSSJ222418.26+002131.6 & 0.409 & 1.97 $\pm$ 0.78 & 5.86 $\pm$ 3.3 & 0.81 $\pm$ 0.06 & $i'$  & 1.34 \\
1237660025032868014 & SDSSJ224901.97-004149.7 & 0.358 & 0.78 $\pm$ 0.19 & 3.16 $\pm$ 0.68 & 0.48 $\pm$ 0.13 & $i'$  & 1.12 \\
1237650804268400981 & SDSSJ101351.61-022106.1 & 0.345 & 1.17 $\pm$ 0.39 & 2.22 $\pm$ 0.34 & 0.66 $\pm$ 0.04 & $g'$ & 1.91
\enddata
\tablecomments{Columns: (1) SDSS object identification; (2) SDSS designation; (3) Redshift; (4) Circularized half-light radius of the single-profile model (R$_{e,c}=\mathrm{R}_e\times \sqrt{b/a}$, where R$_e$ is the half-light radius along the major axis and $b/a$ is the axial ratio); (5) S\'ersic index of the single-profile model; (6) Axial ratio of the single-profile model; (7) CFHT MegaCam filter; (8) Half width at the half maximum of the PSF, in units of kpc at the target's redshift, constructed by combining a set of stars in the CFHT MegaCam image.}
\end{deluxetable*}

\subsection{Color profiles of compact galaxies at $z\sim0.4$\label{color}} 

Three compact BOSS galaxies have multi-band images available in the CFHT MegaCam database. These targets lie at $z\approx0.4$, have similar velocity dispersions of $100<\sigma<140$~km~s$^{-1}$, and straddle the age range between 1 and 2~Gyr.  Two systems - SDSSJ1222+1055 and SDSSJ1234+0818 - exhibit spectra typical for passively evolving galaxies with a more prominent 4000~\AA \ break and weaker Balmer lines (similar to the example spectrum in panel 2 of Figure~\ref{f2}). Our third target with a radial color profile - SDSSJ0207-0618 - shows a spectrum dominated by very strong Balmer lines (similar to the example spectrum in panel 1 of Figure ~\ref{f2}).

\begin{figure*}
\begin{centering}
\includegraphics[scale=0.4]{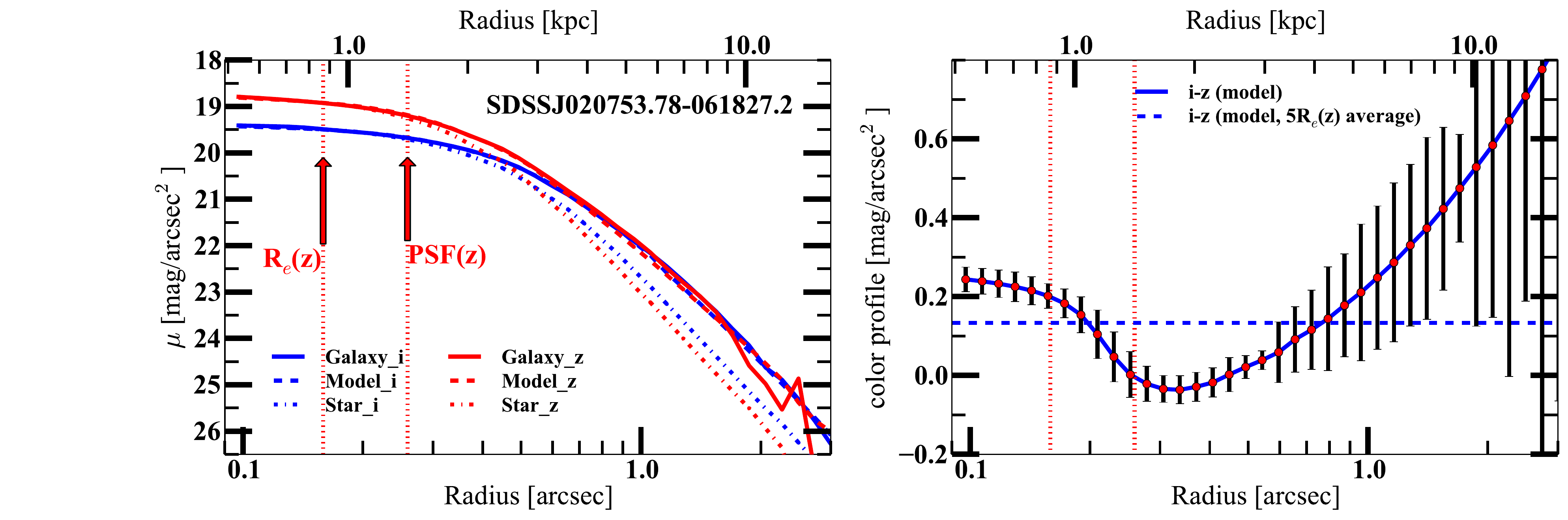}
\includegraphics[scale=0.4]{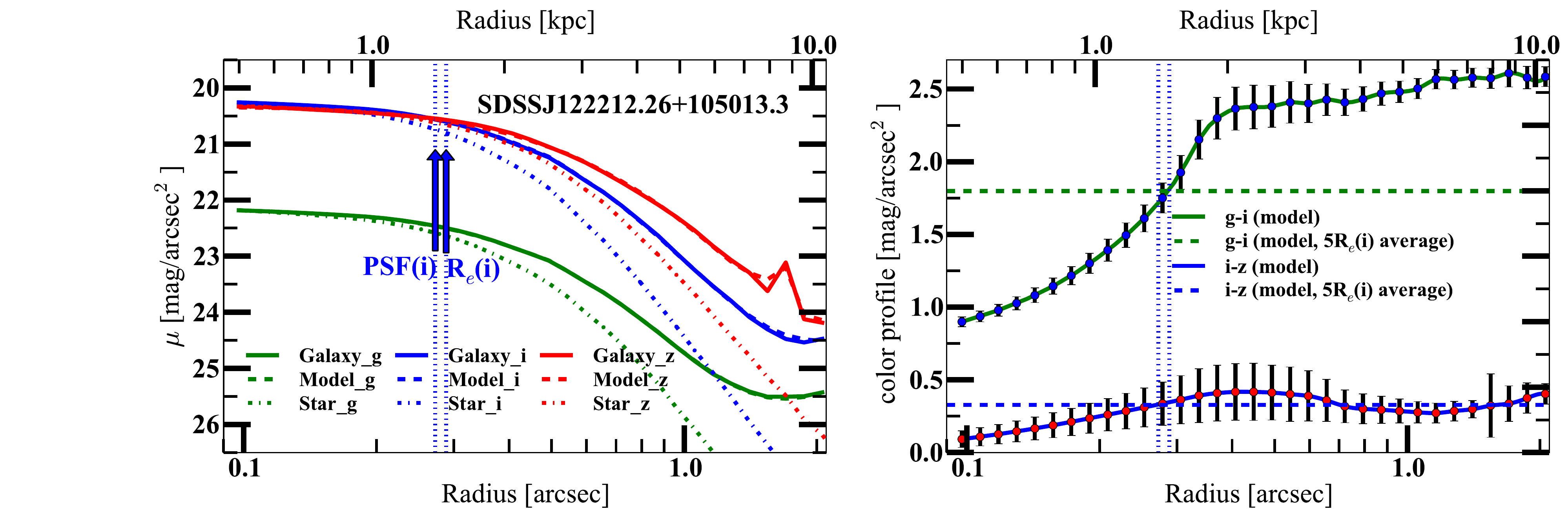}
\includegraphics[scale=0.4]{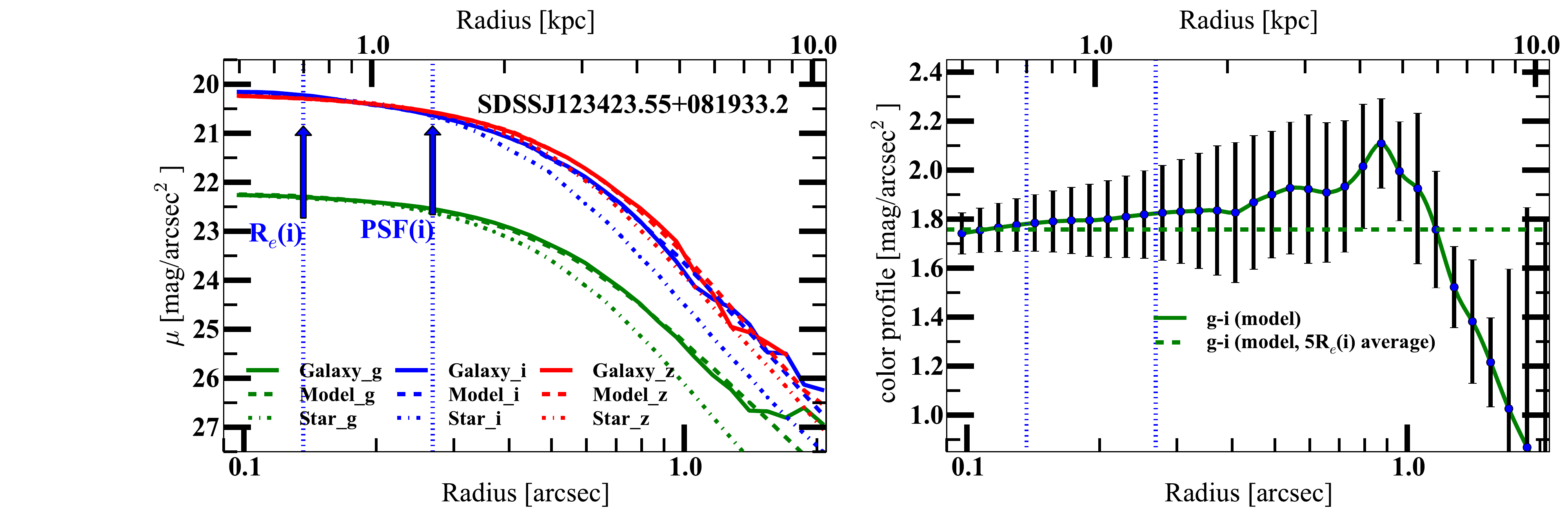}

\caption{Radial profiles of three BOSS intermediate-redshift compacts with available imaging in multiple CFHT MegaCam broad-band filters. Panels on the left-hand side show the observed profile of the galaxy (solid lines), the best fit GALFIT model convolved with the PSF (dashed line), and the profile of the corresponding PSF (dashed-dotted line) in all available MegaCam filters. The half-light radius along major axis and the HWFM(PSF) in the filter with best resolution (used to measure galaxy size reported in Table~\ref{tab2}) are presented with dotted lines and arrows.  Right-hand side panels show radial profiles of galaxy colors constructed from the de-convolved best-fit GALFIT models. The global color value (obtained from total magnitudes provided by GALFIT) is represented with a colored dashed line in each of these three panels. Error bars represent the difference between the observed isophotal surface brightness and the isophotal surface brightness of the model (convolved with PSF). Note that for SDSSJ1234+0819 we show radial profiles in three CFHT MegaCam filters - $g$, $i$, and $z$ - and only one color profile - $g-i$. The object is only marginally resolved in $z$-band image, and the other two broad-band filter images have much better (and comparable) image quality.\label{f4}}
\end{centering}
\end{figure*}

Figure~\ref{f4} shows multi-band radial profiles and related color profiles for three BOSS-CFHT targets. Each left-hand side panel contains the observed surface brightness profiles, best-fit models (convolved with matching PSFs) and the PSF profiles in all available filters. In addition, we use de-convolved best-fit models in different bands to construct observed-frame color profiles (right-hand side panels). The small number of objects in this subsample does not allow us to search for correlations among color gradients and other galaxy properties including stellar age, metallicity, or extinction. However, color profiles based on the best-fit models display changes in color  with respect to the average global value within 5 effective radii (dashed lines in color profile panels of Figure~\ref{f4}) for a large range of galactocentric distances ($0.5\, R_{e}\lesssim r\lesssim10\, R_{e}$). Within one half-light radius there are three different profiles: SDSSJ0207-0618 exhibits a negative $i-z$ color gradient in its central region, SDSSJ1222+1050 has positive color gradients in both $g-i$ and $i-z$, and the $g-i$ profile of SDSSJ1234+0819 is flat within $1\, R_{e}$.

In comparison, using HST imaging, \citet{Gargiulo2012} find a negative central F850LP--F160W color gradient in 70\% of their $1<z<2$ targets and flat central F850LP--F160W  color profiles in the rest of the sample. The dynamical masses of our three targets lie at the lower mass limit of this high redshift sample. Their ages overlap with the ages of 6/11 high-$z$ targets. 

Global color profiles of the three BOSS galaxies follow the trends observed in their central regions. SDSSJ0207-0618 has inner region (within $r\sim1-2\, R_{e}$) that is redder than its outskirts ($r>2\, R_{e}$). This  result agrees with previously reported color profiles of passively evolving galaxies with stellar masses $M_\ast>10^{10}$~M$_\sun$ at $z>1$ \citep{Szomoru2011, Guo2011, Gargiulo2012}. SDSSJ1234+0819 shows an essentially flat profile at all radial scales. In contrast, the most extended and most massive system with the oldest stellar population in our multi-band sample, SDSSJ1222+1050 displays a blue core and red outer regions in $g-i$ color.

Although there are only three compact BOSS systems with available CFHT MegaCam imaging in multiple visible bands, each one shows a different radial dependence in observed colors. Such a small sample of galaxies with similar ages (1-2~Gyr) and gravitational potentials ($\sigma\sim120$km~s$^{-1}$) is not sufficient to constrain the formation and/or evolution of compact galaxies. However, the diversity of observed profiles suggests that these compact systems may form under a wide range of conditions, echoing the conclusions from e.g. \citet{Gargiulo2012}.
      
\section{The relations between structural and dynamical properties of compact galaxies at $z\sim0.4$}\label{fundamental}

\begin{figure*}
\begin{centering}
\includegraphics[scale=0.45]{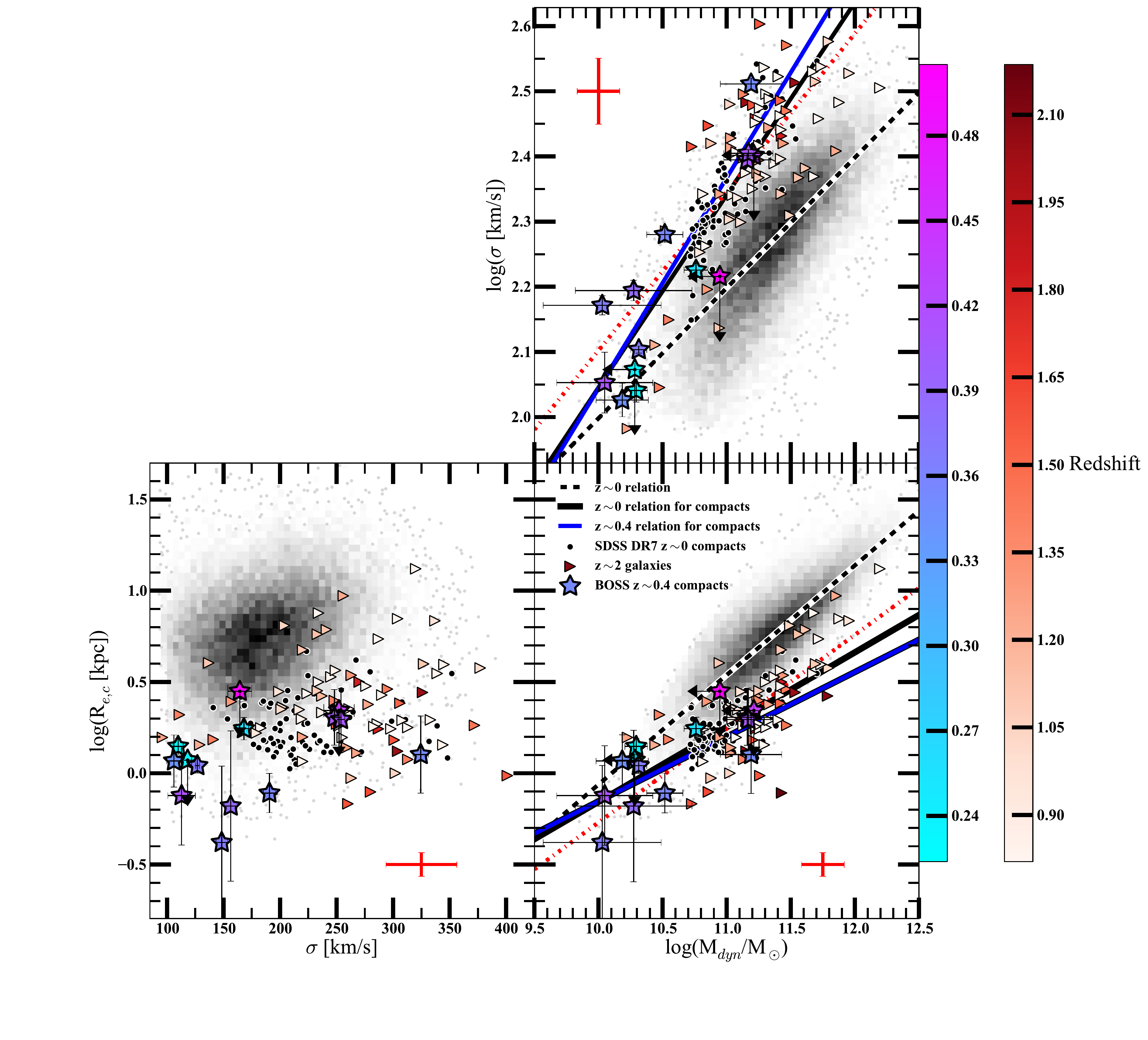}
\caption{Three relations between dynamical and structural properties of BOSS compact galaxies observed with CFHT MegaCam (stars; Table~\ref{tab2}), compared with the properties of massive ellipticals at $z\sim0$ \citep[gray histogram and gray points;][]{Simard2011}, quiescent galaxies at $z>1$ \citep[triangles;][]{vandeSande2013}, and the most compact galaxies found in SDSS at $z<0.2$ \citep[black circles;][] {Trujillo2009, Taylor2010}. The three relations are: circularized half-light radius vs. velocity dispersion ({\it bottom left};  circularized half-light radius vs. dynamical mass ({\it bottom right}); velocity dispersion vs. dynamical mass ({\it top right}). Our BOSS targets at $0.2<z<0.6$ and the high-$z$ comparison sample are color-coded by redshift. Red bars represent average errors for the high-redshift sample. Different lines represent the the best-fit relations: (1) for {\it all} massive quiescent systems at $z\sim0$ and at high $z$ (black and red lines, respectively) and (2) for {\it compact} quiescent systems at $z\sim0$ and at $z\sim0.4$ (gray and blue lines, respectively) in the size - dynamical mass and velocity dispersion - dynamical mass parameter space. We use the size - mass relation to quantify the difference between BOSS compact galaxy sizes and the sizes of $z\sim0$ SDSS passively evolving galaxies at equivalent dynamical masses. We use the velocity dispersion -  dynamical mass relation to infer dynamical masses of BOSS compact targets without available high-resolution images (Section~\ref{select}). \label{f5}} 
\end{centering}
\end{figure*}

To compare measured sizes, S\'ersic indices, and velocity dispersions of the compact BOSS-CFHT galaxies with the structural and dynamical properties of quiescent systems at high redshift and in the nearby Universe, we place the three galaxy samples in the size--velocity dispersion--dynamical mass parameter space (Figure~\ref{f5}). The $z>1$ sample is a spectro-photometric compilation described in \citet{vandeSande2013}. The $0<z<0.2$ galaxies are selected from the \citet{Simard2011} morphological galaxy catalog based on the SDSS DR7. We use spectra of these $z\sim0$ SDSS galaxies to single out quiescent systems with EW$[$O II$]<5$~\AA \ and we re-fit their spectra to obtain the same properties as for the compact BOSS sample at $0.2<z<0.6$ (velocity dispersion, SSP age, and metallicity). In addition, we assemble a set of the most compact systems found among SDSS galaxies \citep{Trujillo2009, Taylor2010}.

Figure~\ref{f5} shows the position of each galaxy sample in the parameter space defined by their structural and dynamical properties \citep[a variation of the dynamical Fundamental Plane,][]{Onorbe2006}. We present: (1) the circularized half-light radius $R_{e,c}$, (2) the stellar velocity dispersion $\sigma$ within a circularized aperture of radius $R_{e,c}$, and (3) the dynamical mass $M_{dyn}$ defined as

\begin{equation} 
M_{dyn}=\frac{\beta\sigma^2 R_{e,c}}{G}, \label{eq:mdyn}
\end{equation}

\noindent where $\beta=5.0\pm0.1$ is a constant scaling factor derived from a comparison between the dynamical mass-to-light ratio ($M/L$) based on the velocity dispersion within one half-radius and $M/L$ from spatially resolved stellar kinematics for a sample of nearby early-type galaxies \citep{Cappellari2006}.


In all three panels of Figure~\ref{f5}, compact BOSS galaxies at $0.2<z<0.6$ (stars) overlap with the most compact objects in the high-$z$ sample (triangles). They also overlap with extreme systems at  $z\sim0$ (black circles). However, the most compact objects in our $z\sim0.4$ sample (with $R_e<0.9$~kpc $[\log(R_e/\mathrm{kpc})<-0.05]$) are up to an order of magnitude smaller than the typical quiescent galaxies at $z\sim0$ at similar velocity dispersions (lower left panel, Figure~\ref{f5}). We note that a very small fraction (0.4\%) of local quiescent systems from the \citet{Simard2011} catalog fall in the region occupied by the most compact massive systems at $z>0$. Visual inspection of SDSS images available for these outliers shows a range of examples: objects near bright stars, clustered systems, objects containing multiple bright clumps, edge-on disks, but also isolated bulge-like systems. Although some of these galaxies may be genuinely dense, their compactness needs to be confirmed through a careful 2D fitting of their surface brightness profiles from better-quality imaging. Compact systems at all redshifts lie away from the locus of typical massive quiescent galaxies in the nearby Universe (grayscale map). 

In the size-dynamical mass parameter space (lower right panel, Figure~\ref{f5}), the best-fit relations for quiescent galaxies at $z\sim0$ and at $z>1$ have similar slopes: $\log(R_{e,c}[$kpc$](z=0))\propto0.63\times\log(M_{dyn}[M_\sun])$ (black dashed line) and  $\log(R_{e,c}\allowbreak[$kpc$](z>1))\propto0.5\times\log(M_{dyn}[M_\sun)])$ (red line). For the compact BOSS galaxies with measured half-light radii the size-dynamical mass relation has a somewhat shallower slope, $\log(R_{e,c}[$kpc$])(z\sim0.4)\propto0.34\times\log(M_{dyn}[M_\sun])$ (blue line), closely resembling the trend that the most compact $z\sim0$ systems follow in size - dynamical mass parameter space (black solid line). However, for all dynamical masses $\gtrsim3\times10^{10}\, M_\sun [\log(M_{dyn}/M_\sun)\gtrsim10.5]$ relations for both BOSS compacts at $0.2<z<0.6$ and high-$z$  objects give galaxy sizes that are $\gtrsim2$ times smaller than the typical size of a $z\sim0$ galaxy at the same dynamical mass. Furthermore, assuming that all of the BOSS compact galaxies follow the same size-dynamical mass relation, their predicted sizes for galaxy masses $>2\times10^{10}M_\sun [\log(M_{dyn}/M_\sun)>10.3]$ are less than half the size of an equally massive ellipticals used as a reference at $z\sim0$ \citep{Shen2003}.\footnote{Here and in Sections~\ref{comp}~and~\ref{model} we assume that the dynamical and stellar masses of compact galaxies are identical because these systems are dominated by stars, as demonstrated by \citet{Conroy2013}.}

Because of their extreme sizes, $M_{dyn}>3\times10^{10}\, M_\sun$ compact systems at all redshifts have lower dynamical masses than typical quiescent galaxies at $z\sim0$ with the same velocity dispersion (upper panel, Figure~\ref{f5}). Local and high-$z$ velocity dispersions of quiescent massive galaxies are related to dynamical mass with similar slopes: $\log\left(\sigma [\mathrm{km\, s}^{-1}]\left(z\sim0\right)\right)\propto(0.207\pm0.001)\times\log\left(M_{dyn}[M_\sun]\right)$, $\allowbreak\log\left(\sigma [\mathrm{km\, s}^{-1}]\left(z>1\right)\right)\propto(0.24\pm0.3)\times\log\left(M_{dyn}[M_\sun]\right)$. The subsample of compact BOSS galaxies follows the relation: 

\begin{equation}\label{eq: sigma}
\begin{split}
\log\left(\sigma [\mathrm{km\, s}^{-1}]\left(z\sim0.4\right)\right)&=(0.32\pm0.06)\times\log\left(M_{dyn}[M_\sun]\right) \\
                                                                                                            & -(1.2\pm0.6).  
\end{split}
\end{equation}  

\noindent The slope in  Eq.~\ref{eq: sigma} is steeper than the trend defined by typical massive systems at $z\sim0$. Conversely, velocity dispersions of compact galaxies at $0.2<z<0.6$ and at $z\sim0$ follow the same trend with dynamical mass. Although the slope of the intermediate-redshift velocity dispersion - dynamical mass relation for compact systems is somewhat steeper than the one for all quiescent galaxies at high-$z$, the two $d(\log\sigma)/d(\log M_{dyn})$ values are consistent within the fitting errors. We use Eq.~\ref{eq: sigma} to infer dynamical masses for the rest of the compact intermediate-redshift galaxies with velocity dispersions measured from their BOSS spectra, but without available high-quality imaging (Section~\ref{select}).  

Structural properties of a subset of BOSS compact systems combined with measured stellar kinematics confirm that in the size - velocity dispersion - dynamical mass parameter space they occupy the region populated by the densest galaxies at $z\sim0$ and $z>1$. In contrast, at $z\sim0$ typical quiescent galaxies with similar velocity dispersions have sizes and dynamical masses several times larger. 

Another argument that the BOSS sample must be compact is a statistical one. We draw 14 compact objects from our sample of~$\sim200$ candidates. If we assume that these objects are a random subsample, a simple resampling exercise shows that in order to draw a subsample of 14 compact galaxies from a parent sample of ~200 objects at the 90\% confidence level, at least 82\% of the parent sample must be compact. Thus we conclude that, by selecting BOSS point sources with the spectra of redshifted galaxies, we construct a sample of compact galaxies that can be further used to estimate the number density of these systems in the redshift range $0.2<z<0.6$.    

\section{Number density of intermediate redshift compact galaxies}\label{nd}

The large sample of uniformly selected compact system candidates  allows us to trace their number density in narrow bins within the redshift range covered by the sample. Here we describe a procedure that we use to compute number densities of intermediate-redshift BOSS compact galaxies. We compare these lower limits to the reported abundances of similarly compact massive passively evolving galaxies at lower and higher redshift. Finally, we place the number density of intermediate-redshift compacts in the context of the semi-analytic models of massive galaxy evolution.     

\subsection{The Method\label{spec}}

\begin{figure*}
\begin{centering}
\includegraphics[scale=0.35]{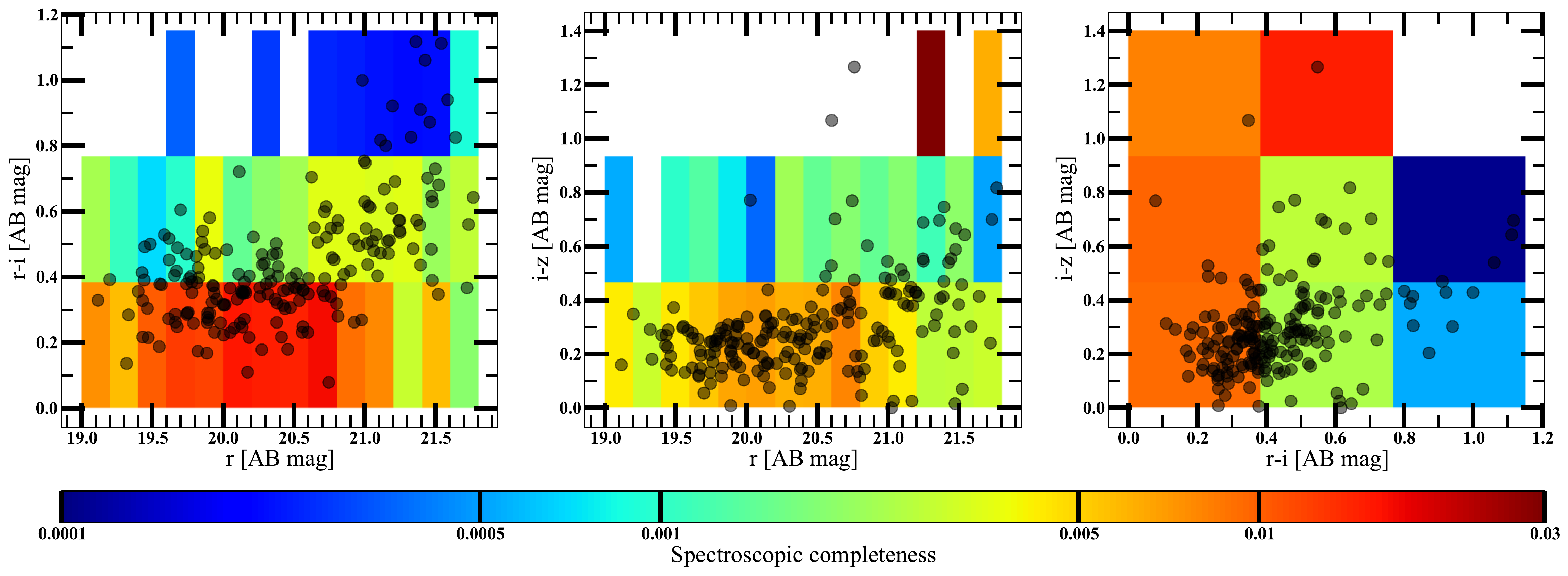}
\caption{Three-dimensional fractional spectroscopic completeness of the BOSS survey in the $r-i$ color, $i-z$ color, and $r$~magnitude range of the identified intermediate-redshift BOSS compact galaxies. The panels show three projections of the parameter space we use to calculate this fractional spectroscopic completeness. The cell size in each panel is four times larger than the average photometric error for BOSS compact galaxies in the corresponding color or magnitude. Cells are color-coded by the fractional spectroscopic completeness, i.e. by the ratio between the number of SDSS photometric point sources and the number of BOSS spectra in the $r-i$ color , $i-z$ color, and $r$ magnitude range defined by that cell. Gray circles show the positions of BOSS compact galaxies in all three projections of the color-color-magnitude parameter space. In practice, we use a cell centered on the position of an individual BOSS target in this parameter space to estimate corresponding spectroscopic completeness factor for each compact galaxy in our sample. \label{f6}}
\end{centering}
\end{figure*}

To calculate the number density of BOSS compact galaxy candidates we begin with the assumption that these intermediate-redshift compact galaxies are a random subsample of the whole population that BOSS could survey. Thus we correct only for spectroscopic (in)completeness of the BOSS survey within the magnitude and color range where our targets are found.

We use the  distribution of compact system candidates from our sample in a three-dimensional parameter space: $r$ magnitude, $r-i$ color, and $i-z$ color. We use the observed sample to define a range of values for each parameter:  (1) $19 < r < 21.8$, (2) $0.05 < r-i < 1.15$, and (3) $0 < i-z < 1.4$. 

From the SDSS photometric database we then select all point sources in two representative 50 sq. degrees fields at Galactic latitudes $b\sim65\arcdeg$ and $b\sim-45\arcdeg$ that fall within these photometric boundaries. These fields lie at the median northern and southern Galactic latitudes of the BOSS coverage. We use the same fields to select all spectroscopic targets within the same color and color-magnitude ranges. In this parameter space we then define a grid of cells  with size $\sim4$ times larger than the absolute average error in the corresponding colors and magnitude of the compact galaxy sample. The spectroscopic completeness factor for each compact target is then a ratio between the number of  BOSS spectroscopic targets and the number of  SDSS point sources in a cell centered at the position of that target in this three-dimensional grid.

Figure~\ref{f6} illustrates this concept. Each panel is one projection of the parameter space used for estimating the spectroscopic completeness. The size of each cell - $\Delta(r)\sim0.2$~mag, $\Delta(r-i)\sim0.4$~mag, $\Delta(i-z)\sim0.5$~mag - is related to the average photometric error for the intermediate-redshift compact BOSS systems. Gray circles display the positions of our targets in color-magnitude and color-color diagrams. The only difference between the representation in Figure~\ref{f6} and our calculations is that in practice each compact galaxy has a unique three-dimensional cell centered at the colors and $r$ magnitude of that galaxy. This approach allows us to estimate the spectroscopic completeness for all but 2 objects in our sample of $\sim200$ galaxies. These two galaxies have much redder $i-z$ color than the typical SDSS photometric point sources of the same $r$ magnitude. 

By selecting our compact candidates from SDSS/BOSS, the largest photo-spectroscopic extragalactic survey to date, we minimize a major limitation of galaxy number density estimates: cosmic variance. Another limiting factor in measurements based on a single survey with multi-layered selection algorithm, such as BOSS quasar survey \citep{Ross2012}, are systematic uncertainties. Comparison with other independent surveys with well-understood selection criteria is the only approach that can provide calibration for this type of errors. We plan to address this important question in our future investigations (Damjanov et al. 2014, in prep).

Cells in Figure~\ref{f6} are color-coded by the spectroscopic completeness; very low fractions shown as color bar labels demonstrate two important properties of the BOSS quasar survey: (1) the survey is sparse \citep[e.g.,][]{Eisenstein2011}, and (2) by searching for $2.2<z<3.5$ quasars preferentially in the $0<g-r<0.5$ color range \citep[e.g.,][]{Ross2012}, the survey selects against red compact galaxies at $0.2<z<0.6$. Thus the BOSS quasar survey is not optimized for recovering a representative sample of intermediate-redshift compact quiescent systems. However, this survey provides a hard lower limit on the number density of these extreme systems, as we describe below. 

\subsection{The Number Density Estimates}\label{res}

After applying the  correction for BOSS spectroscopic completeness, we calculate the number densities of compact galaxies in $\Delta(z)=0.1$ redshift bins within the $0.2<z<0.6$ range by dividing the corrected numbers of compact galaxy sources by the volumes spanned by the 6373.2 sq. degree BOSS survey \citep{Ahn2014} in each redshift bin. Because $\gtrsim82\%$ of all galaxies in our sample have to be compact to provide a subsample of 14 compact objects with measured structural properties (Section~\ref{fundamental}), we draw 1000 unique subsamples of the 162 compact BOSS galaxies  (representing 82\% of the total number of compacts in our sample) and derive a galaxy number density corrected for spectroscopic completeness for each of them. The final number density of compact galaxies in each redshift bin is the mean value of these 1000 number densities weighted by the Poisson errors associated with each subsample. We repeat the analysis for different lower limits in dynamical mass (obtained from the measured velocity dispersions; Eq.~\ref{eq: sigma}) and in galaxy age (i.e. the age of the best-fit SSP model from Table~\ref{tab1} transformed into the formation redshift). 

During this procedure we propagate the errors in the parameters of the best-fitting velocity dispersion - dynamical mass relation (Section~\ref{fundamental}) by selecting a range of velocity dispersion thresholds for each lower limit in dynamical mass. Our simulations also include a range of color and magnitude bin sizes and two different sets of colors - ($g-r$, $r-i$) and ($r-i$,$i-z$) - that we use to estimate the spectroscopic completeness of BOSS (Section~\ref{spec}). These variations in the combination of colors and bin sizes produce additional scatter in the resulting galaxy number densities. We add this uncertainty to the total error budget and compare our results with number densities of compact systems reported at $z\sim0$ and high redshift (Section~\ref{comp}).

\begin{deluxetable*}{cccc}
\tabletypesize{\scriptsize}
\tablecaption{Number densities of intermediate-redshift compact galaxies selected from BOSS \label{tab3}}
\tablewidth{0pt}
\tablehead{
\colhead{Redshift range} & \colhead{Mass cut} & \colhead{Formation redshift} & \colhead{Number density\tablenotemark{a}} \\
\colhead{} & \colhead{[$\times10^{10}\, M_\sun$]} & \colhead{} & \colhead{[Mpc$^{-3}$]}  \\
\colhead{(1)} & \colhead{(2)} & \colhead{(3)} & \colhead{(4)} 
}
\startdata
$0.2<z<0.3$ & $>1$ & Any & $ (1.6^{+0.4}_{-0.4})\times10^{-5}$\\ 
$0.3<z<0.4$ & $>1$ & Any & $ (1.07^{+0.25}_{-0.25})\times10^{-5}$\\
$0.4<z<0.5$ & $>1$ & Any & $ (6.7^{+2.2}_{-2.2})\times10^{-6}$\\
$0.5<z<0.6$ & $>1$ & Any & $ (5.1^{+1.8}_{-1.8})\times10^{-6}$\\ \hline
$0.2<z<0.3$ & $>3$ & Any & $ (1.2^{+0.4}_{-0.5})\times10^{-5}$\\ 
$0.3<z<0.4$ & $>3$ & Any & $ (7.8^{+2.5}_{-3.1})\times10^{-6}$\\
$0.4<z<0.5$ & $>3$ & Any & $ (5.6^{+1.8}_{-2.3})\times10^{-6}$\\
$0.5<z<0.6$ & $>3$ & Any & $ (5.0^{+1.9}_{-1.8})\times10^{-6}$\\ \hline
$0.2<z<0.3$ & $>5$ & Any & $ (8.1^{+4.2}_{-4.8})\times10^{-6}$\\ 
$0.3<z<0.4$ & $>5$ & Any & $ (5.9^{+2.4}_{-2.6})\times10^{-6}$\\
$0.4<z<0.5$ & $>5$ & Any & $ (4.6^{+2.0}_{-1.8})\times10^{-6}$\\
$0.5<z<0.6$ & $>5$ & Any & $ (4.7^{+1.9}_{-1.6})\times10^{-6}$\\ \hline
$0.2<z<0.3$ & $>8$ & Any & $ (4.1^{+4.5}_{-3.0})\times10^{-6}$\\ 
$0.3<z<0.4$ & $>8$ & Any & $ (2.8^{+3.6}_{-2.0})\times10^{-6}$\\
$0.4<z<0.5$ & $>8$ & Any & $ (2.7^{+2.8}_{-2.1})\times10^{-6}$\\
$0.5<z<0.6$ & $>8$ & Any & $ (4.8^{+1.4}_{-3.4})\times10^{-6}$\\ \hline
$0.2<z<0.3$ & $>10$ & Any & $ (2.6^{+4.0}_{-2.5})\times10^{-6}$\\ 
$0.3<z<0.4$ & $>10$ & Any & $ (1.7^{+3.3}_{-1.7})\times10^{-6}$\\
$0.4<z<0.5$ & $>10$ & Any & $ (1.9^{+3.4}_{-1.7})\times10^{-6}$\\
$0.5<z<0.6$ & $>10$ & Any & $ (4.1^{+2.4}_{-1.7})\times10^{-6}$\\ \hline\hline
$0.2<z<0.3$\tablenotemark{b} & $>6$ & $>2$ & $ (2.9^{+1.8}_{-1.4})\times10^{-6}$\\\ 
$0.3<z<0.4$\tablenotemark{b} & $>6$ & $>2$ & $ (1.5^{+0.8}_{-1.3})\times10^{-6}$\\
$0.4<z<0.5$\tablenotemark{b} & $>6$ & $>2$ & $ (1.3^{+0.7}_{-1.0})\times10^{-6}$\\
$0.5<z<0.6$\tablenotemark{b} & $>6$ & $>2$ & $ (8.7^{+4.8}_{-5.5})\times10^{-7}$\\ \hline
$0.2<z<0.3$\tablenotemark{c} & $>8$ & $>2$ & $ (2.6^{+1.2}_{-2.4})\times10^{-6}$\\ 
$0.3<z<0.4$\tablenotemark{c} & $>8$ & $>2$ & $ (1.0^{+1.0}_{-0.7})\times10^{-6}$\\
$0.4<z<0.5$\tablenotemark{c} & $>8$ & $>2$ & $ (9.0^{+9.0}_{-8.4})\times10^{-7}$\\
$0.5<z<0.6$\tablenotemark{c} & $>8$ & $>2$ & $ (8.3^{+5.2}_{-5.2})\times10^{-7}$
\enddata
\tablecomments{Number densities above the double horizontal line are presented in the left panel of Figure~\ref{f7} and can be compared to number densities at lower and higher redshifts (Section~\ref{comp}). Number densities below double horizontal line are compared to the predictions of semi-analytic models in Section~\ref{model} and in the right panel of Figure~\ref{f7}.}
\tablenotetext{a}{{\it Lower limit} on number density of compact massive galaxies in these intermediate-redshift bins.}
\tablenotetext{b}{To be compared to the expected number densities of massive galaxy relics based on semi-analytical models under the assumption that the fractional mass growth of these system over the redshift range $0<z<2$ is $\frac{\Delta M}{M}<0.3$ (Quillis \& Trujillo 2013).}
 \tablenotetext{c}{To be compared to the expected number densities of massive galaxy relics based on semi-analytical models under the assumption that the fractional mass growth of these system over the redshift range $0<z<2$ is $\frac{\Delta M}{M}<0.1$ (Quillis \& Trujillo 2013).}
\end{deluxetable*}

Our estimates of number densities are lower limits on the abundance of compact systems in the redshift range $0.2<z<0.6$ because the photometric selection algorithm of the BOSS quasar survey \citep{Ross2012} was not designed for tracing massive galaxies across intermediate redshift range. Low spectroscopic completeness of BOSS within color and magnitude ranges where our candidates are found, shown in Figure~\ref{f6}, confirms that these systems represent tails of the distribution of BOSS spectroscopic targets. These selection effects force the exclusion of a large fraction of the redder, and thus more massive and/or more distant, compact candidates from our sample (Damjanov et al. 2014, in prep). Table~\ref{tab3} shows the resulting number densities for different combinations of galaxy redshift, dynamical mass, and age cuts. 

\subsection{Comparison with number densities of compact quiescent galaxies in the nearby and distant Universe}\label{comp}

\begin{figure*}
\begin{centering}
\includegraphics[scale=0.25]{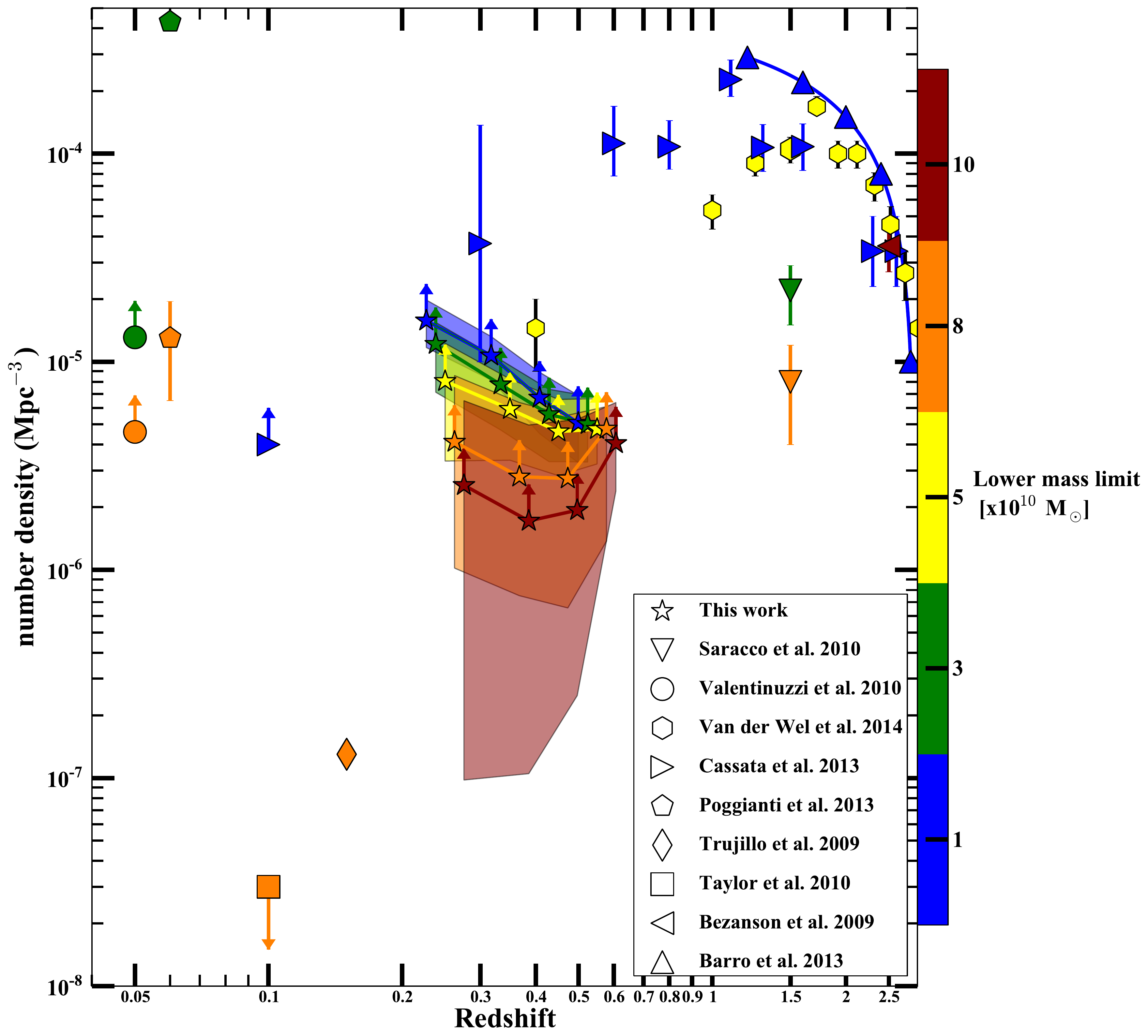}
\includegraphics[scale=0.25]{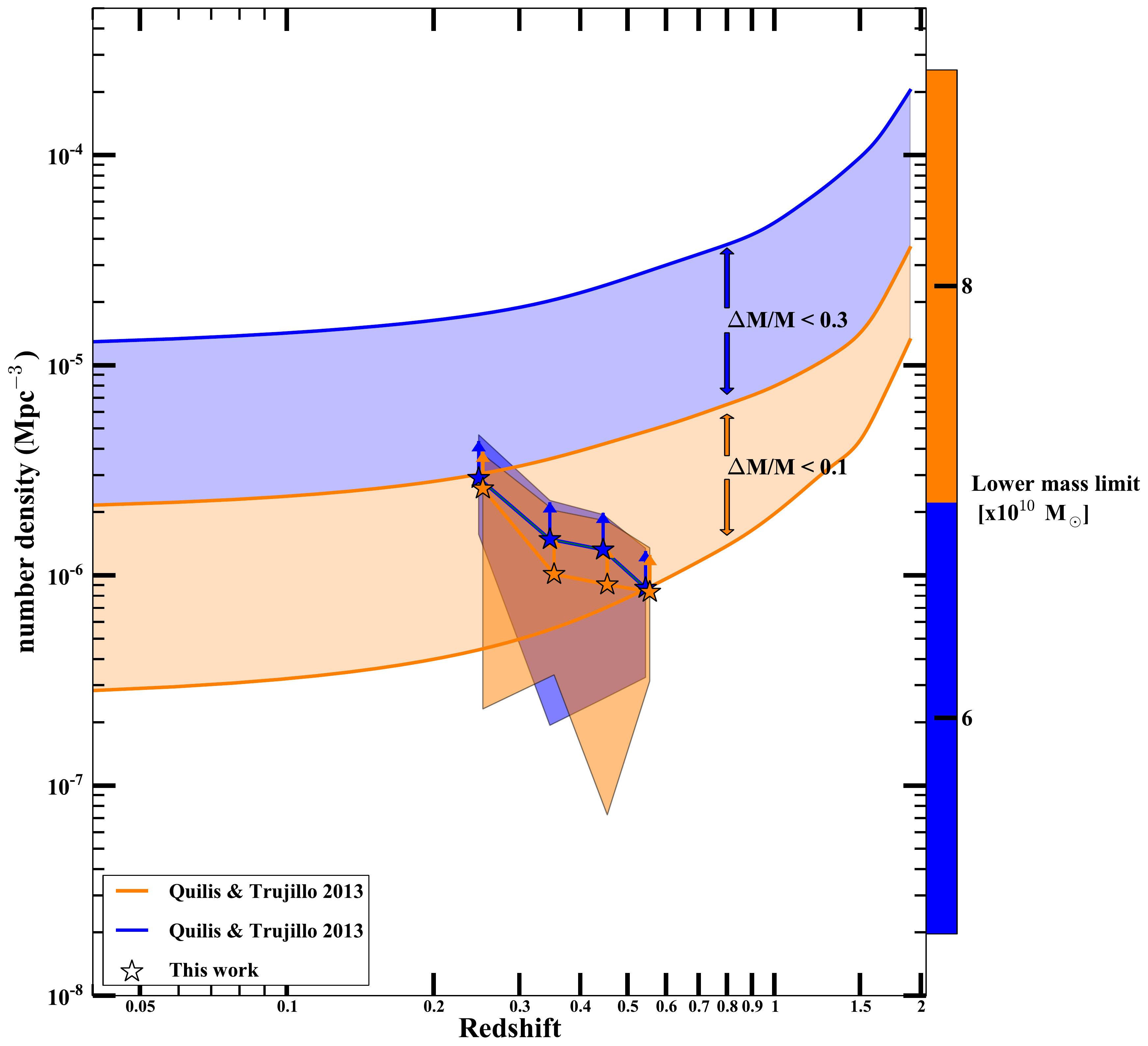}

\caption{The number density of compact quiescent galaxies as a function of redshift.  {\it Left panel:} Comparison with measured abundance of compact massive quiescent galaxies at high and low redshift. The number densities we derive for the redshift range $0.2<z<0.6$ (stars and filled regions denoting $\pm1\sigma$ ranges) are the lower limits on the abundance of massive compact systems in this redshift range.  Different symbols correspond to different studies that we compare with our results based on the BOSS compact galaxy sample. All points are color-coded according to the minimum galaxy mass of the sample used to calculate galaxy number density. Although comparison with the results from literature is not straightforward, the BOSS sample of compact quiescent galaxies at $0.2<z<0.6$ provides intermediate-redshift galaxy number densities that are consistent with majority of values reported at other redshifts. {\it Right panel:} comparison with semi-analytic model estimates. The filled regions represent the predictions of the semi-analytic models for two trends of growth for massive quiescent galaxies formed at $z>2$ \citep{Quilis2013}: (1) galaxies can acquire up to 10\% of their mass from $z=2$ to $z=0$ (orange band) or (2) galaxies can gain up to $30\%$ of their mass in this redshift range (blue band). The number densities of the most massive systems in our $0.2<z<0.6$ BOSS sample that were formed at redshift $z>2$ are shown as stars and filled 
$\pm1\sigma$ regions. The points are color-coded according to the lower galaxy mass threshold. The abundances of the oldest and most massive BOSS quiescent compacts follow the predictions of semi-analytic models. See text for details.\label{f7}}
\end{centering}
\end{figure*}

Evolution of  the number density of compact massive passively evolving galaxies over the redshift range  $0<z<3$ is a crucially important constraint on the models of massive galaxy assembly \citep[e.g.,][]{Hopkins2009, Shankar2012}. In recent years several groups have worked to constrain this important quantity at $z\sim0$ \citep[e.g.,][]{Trujillo2009, Taylor2010, Valentinuzzi2010, Poggianti2013} and $z\gtrsim1$ \citep[e.g.,][]{Bezanson2009, Saracco2010, Cassata2011, Cassata2013, Barro2013, Stefanon2013}. However, the range between $z\sim0.2$ and $z\sim0.6$ has not been extensively explored \citep{Cassata2013}.  Thus the number densities of compact BOSS galaxies at intermediate redshift that we report here are a valuable link between these extreme systems in two vastly different redshift regimes. 

In Figure ~\ref{f7} we trace the abundance of compact massive galaxies over $\sim11$~Gyr of cosmic time (left panel) and compare the number densities of massive old systems among our BOSS targets with the predictions of several semi-analytic models (right panel). Left-hand side panel shows different symbols, corresponding to different studies in this redshift range, that are color-coded by the lower mass limit of each sample. We show results in four redshift bins spanning the redshift range $0.2<z<0.6$ (stars). We highlight different mass cuts in order to compare abundances of the compact BOSS systems with the low- and high-redshift values taken from the literature. Comparison between our results and previously reported number densities at $0<z<2.7$ are not straightforward because each study represented in Figure~\ref{f7} uses (a) different definitions for galaxy compactness and (b) different cuts in galaxy mass and/or age to select a representative sample.\footnote{Number densities for all comparison samples are computed using the same (standard) cosmological model parameters that we use in this analysis and list in Section~\ref{intro}.}

At $z<0.2$ it is not clear how rare compact massive systems really are. Different studies reach different conclusions. \citet{Trujillo2009} search for massive ($M_\ast>8\times10^{10}\, M_\sun$) galaxies in SDSS  DR6 with $R_e<1.5$~kpc and find a very small fraction, 0.03\%, that translates into a number density of massive compact galaxies at $\sim0.15$ of $n=1.3\times10^{-7}$~Mpc$^{-3}$ (orange diamond). Galaxies in this sample are predominantly young quiescent systems that could not have formed at very high redshift. The lower limit on the number density of similarly massive compact BOSS systems at $z\sim0.25$ is $\sim30$ times higher (Table~\ref{tab3}, orange stars connected by solid line in the left panel of Figure~\ref{f7}). If we select only massive BOSS targets with sizes $<1.5$~kpc (based on Eq.~\ref{eq: sigma} and measured velocity dispersion), the number density in our lowest redshift bin ($z\sim0.25$) becomes to $n=(1.7\pm0.5)\times 10^{-6}$~Mpc$^{-3}$ , still an order of magnitude above the \citet{Trujillo2009} $z\sim0.15$ result.

In the left panel of Figure~\ref{f7} we also show the number density of red massive ($M_\ast>8\times10^{10}\, M_\sun$) compact SDSS DR7 galaxies at $z\sim0$ from \citet[][orange square]{Taylor2010}. All systems in this sample are more than two times smaller than normal massive galaxies \citep{Shen2003}. This definition of compactness holds for $\gtrsim90\%$ of our sample. In addition \citep[and unlike][]{Trujillo2009}, \citet{Taylor2010} use a color cut to select only red (old) objects with k-corrected colors $^{0.1}(u-r)>2.5$. Thus a significant fraction of this sample formed at $z\gtrsim2$. We apply the same selection criterion to select red objects among our targets and obtain a number density at $z\sim0.25$ of $n=(1.6\pm0.9)\times10^{-6}$~Mpc$^{-3}$, again $20-80$ times higher than the abundance of red compacts at $z\sim0.1$. The number densities of intermediate-redshift compacts in BOSS selected using either criteria from \citet{Trujillo2009} or from \citet{Taylor2010} fall close to our values for the highest mass threshold in the left panel of Figure~\ref{f7} (red line connecting red stars).

On the other end of the spectrum of reported number densities for local compacts are the results of the WINGS \citep{Valentinuzzi2010} and the PM2GC \citep{Poggianti2013} surveys. Compact galaxies in the WINGS cluster survey have masses $M_\ast>3\times10^{10}\, M_\sun$ and surface mass densities within the half-radius $\Sigma_{50}=\frac{0.5M_\ast}{R_{e}^2\pi}\geqslant3\times10^9\, M_\sun$~kpc$^{-2}$. All BOSS compact targets with $M_{dyn}>3\times10^{10}\, M_\sun$ have mass surface densities above this threshold. Thus in the left panel of Figure~\ref{f7} we can directly compare our intermediate-redshift number densities (green/orange line connecting green/orange stars) to the corresponding values at $z\sim0$ from \citet[][green/orange circles]{Valentinuzzi2010}. We find excellent agreement. 

The only caveat in our comparison with \citet{Valentinuzzi2010} is that they assume that compact galaxies do not reside outside of clusters (i.e., they use the total comoving volume up to the redshifts of their clusters, instead of the fraction covered by the survey, to calculate compact galaxy number density). Based on visual inspection of the SDSS and CFHT MegaCam images, our BOSS compacts are both in groups/clusters and in isolation.  

\citet{Poggianti2013} report an abundance of local compact systems for the general field (green and orange pentagons). The compactness of these objects is defined by the same threshold as in \citet{Valentinuzzi2010}. This number density for galaxy masses $M_\ast>3\times10^{10}\, M_\sun$ is $\sim40$ times larger than the number density of similarly massive BOSS compacts in our lowest redshift bin at $z\sim0.25$. When the selection criteria from \citet{Trujillo2009} are applied, the number density of $M_\ast>8\times10^{10}\, M_\sun$, $R_e<1.5$~kpc systems in the PM2GC becomes an order of magnitude lower. However, it is still $\sim10$ times higher than the number density of BOSS compacts at $z\sim0.25$ selected in the same fashion. Although the selection effect discussed in Section~\ref{res} define the number densities we present here as lower limits, it is unlikely that the total exceeds our estimate by a very large factor. 

In the redshift range of our analysis, \citet{Cassata2013} report on the number density of compact and ultra-compact massive quiescent galaxies based on the spectro-photometric data of the GOODS survey (right pointing triangles). In this analysis ultra-compact systems are defined to be more than 2.5 times smaller than the passively evolving SDSS galaxies of the same stellar mass. If we compare the dynamical mass - size relation for BOSS compact galaxies with the upper limit on the size of ultra-compact galaxies as a function of size from \citet{Cassata2013}, the $\Delta\log(R_e)<-0.4$~dex requirement is fulfilled by our targets at dynamical mass of $M_{dyn}\gtrsim7\times10^{10}\, M_\sun$. The lower limit on the number density of these massive BOSS compacts in the redshift range $0.2<z<0.4$ is $n=(5\pm0.9)\times10^{-6}$~Mpc$^{-3}$, $\sim0.2$~dex below the $\pm1\sigma$ range reported in \citet{Cassata2013}. Formally, these two results agree.        
           
Very recent results of the CANDELS and the 3D-HST surveys \citep{vanderWel2014} show strong evolution in number density of the most compact early-type systems (i.e., quiescent based on the rest-frame $UVJ$ color selection). In this study half-light radii are not circularized and the compactness criterium (for systems with stellar masses $M_\ast > 5\times10^{10}\, M_\sun$) takes into account the slope of observed size - mass relation: $R_e/(M_\ast/10^{11}\, M_\sun)^{0.75}<2.5$~kpc. Sizes of all BOSS compact targets with dynamical masses above the CANDELS+3D-HST sample threshold (yellow stars in the left panel of Figure~\ref{f7}), corrected to correspond to the radii along major galaxy axes, pass this compactness test.  Thus the left panel of Figure~\ref{f7} displays that the only compact galaxy number density at $z<1$ for the CANDELS+3D-HST sample (yellow hexagon at $z=0.4$) lends support to the abundances of BOSS compacts we find in the redshift range $0.2<z<0.6$.
           
Massive compact galaxies constitute a significant fraction of massive passively evolving systems at high redshift \citep[e.g,][]{Cassata2011, Cassata2013}.  The left-hand side of Figure~\ref{f7} shows that number densities of $M_{dyn}>3\times10^{10}\, M_\sun$ (green star) and $M_{dyn}>8\times10^{10}\, M_\sun$ (orange star) in the highest redshift bin of our range ($0.5<z<0.6$) agree with the estimates for similarly massive and compact systems at $z\sim1.5$ \citep[][inverted triangles]{Saracco2010}.  We compute the largest correction factor for our sample, its spectroscopic completeness (Figure~\ref{f6}). The complexity of the selection criteria of the parent BOSS sample of quasar candidates prevents us from including other, lower-order effects. However, any of these corrections would only increase the number densities that we provide here, thus increasing the level of agreement with high-$z$ points. 

Only a fraction of these intermediate massive compacts are old enough to be relics of dense high-z systems (Section~\ref{select}). Number densities of $0.6<z<2.5$ ultra-compact systems from \citet[][right pointing triangles]{Cassata2013}, of $1<z<2.5$ compact galaxies from \citet{Barro2013} with $\frac{M_\ast}{R_e^{1.5}}>10.3\, M_\sun$~kpc$^{-1.5}$ (triangles), of $z\sim2.5$ quiescent systems with stellar mass densities $\rho_{50}>10^9\, M_\sun$~kpc$^{-3}$ from \citet[][left pointing triangle]{Bezanson2009}, and of $z>1$ compact early-types from \citet[][hexagons]{vanderWel2014} are consistent  with our lower limits, even though these high-$z$ values are up to two orders of magnitude higher than the lower limit abundances at $z\sim0.55$. 

\subsection{Comparison with models}\label{model}

Finally, we compare number densities of the BOSS intermediate-redshift  compacts formed at $z\gtrsim2$ with the predictions of three semi-analytical models based on the Millennium simulation \citep{Quilis2013}. Orange and blue tracks in the right panel of Figure~\ref{f7} show the evolution in number density of massive galaxies that form at $z\gtrsim2$, gain less than 10\% or 30\% in stellar mass from $z=2$ and $z=0$ through minor mergers, and end up as $M_\ast>8\times10^{10}\, M_\sun$ quiescent systems at $z=0$. Abundances of equally old compact BOSS galaxies at $0.2<z<0.6$ with dynamical masses that are 70\% and 100\% of the threshold model mass at $z=0$ are given in the last rows of  Table~\ref{tab3}. The right-hand  panel of Figure~\ref{f7} shows these number densities as stars, color-coded according to the lower galaxy mass threshold, encompassed by $\pm1\sigma$ error regions (shaded areas). The lower mass limit (that translates into $M_{dyn}>6\times10^{10}\, M_\sun$) imposes an extreme requirement that all of the $\frac{\Delta M}{M}\leqslant0.3$ mass growth happens at $0<z<0.6$. The upper mass limit of $M_{dyn}>8\times10^{10}\, M_\sun$ allows minimum mass growth in this redshift range. Interestingly, our lower limits on number densities are consistent with the predicted evolutionary track for massive systems that acquire less than $10\%$ of their mass since $z\sim2$.

\section{Summary}

Large spectro-photometric databases such as SDSS/BOSS are an ideal starting point for tracing galaxies that are common in distant Universe ($z\gtrsim1$) but become rare with decreasing redshift. We use BOSS spectroscopic data on point-like photometric sources in SDSS to construct a sample of $\sim200$ compact galaxy candidates  at  $0.2<z<0.6$ with spectral features typical for quiescent systems. Taking advantage of publicly available CFHT MegaCam broad-band images in the visible wavelength regime, we measure sizes and other structural properties for a subset of 14 compact candidates from BOSS. 

All of BOSS-CFHT targets are indeed compact and a simple statistical argument implies that at least 82\% of the parent BOSS sample (or $\sim160$ galaxies) have to be compact too. We compare our intermediate redshift sample with spectro-photometric datasets of massive quiescent galaxies populating other redshift regimes. We conclude:       
      
\begin{itemize}

\item Based on their spectra, BOSS compact galaxy candidates span a range in velocity dispersion between 100~km s$^{-1}$ and 320~km s$^{-1}$. Fewer than $10\%$ of the sample are extremely young quiescent systems in a post-starburst phase ($t_{age}\lesssim500$~Myr), $50\%$ of the sample are E+A galaxies ($500$~Myr$<t_{age}<2$~Gyr), and 20\% of the targets are dominated by very old stellar population placing their formation redshift at $z_{form}>2$ (Figure~\ref{f1}).   
\item 2D fitting of the light profiles for a subset of 14 of BOSS  candidates reveals that all are extreme objects with small sizes, very similar to (or smaller than) the high-redshift compact systems and the most compact systems found at $z\sim0$. Our targets show a large offset from the locus of typical massive $z\sim0$ galaxies in the size - velocity dispersion parameter space (Figure~\ref{f5}).
\item We convert measured structural and dynamical parameters for the BOSS-CFHT subsample into dynamical masses. The size - dynamical mass relation for intermediate-redshift compact galaxies lies below the $z\sim0$ relation for typical massive galaxies with dynamical masses $M_{dyn}>10^{10}\, M_\sun$. Furthermore, all  $M_{dyn}>3\times10^{10}\, M_\sun$ systems following our size -dynamical mass relation are less than half the size of equally massive systems following the $z\sim0$ relation.  In the velocity dispersion - dynamical mass parameter space our targets occupy the region of higher velocity dispersions than their massive local counterparts for all masses above $3\times10^{10}\, M_\sun$. 

\end{itemize}      

Completeness-corrected numbers of BOSS compacts in redshift bins of $\Delta z=0.1$ provide robust lower limits on the number densities of the most compact massive galaxies at $0.2<z<0.6$. We compare these abundances with similar systems in different redshift regimes and with theoretical predictions (Figure~\ref{f7}). We find:

 \begin{itemize}
   
\item Compact galaxy number densities at intermediate redshift are lower than but formally consistent with the number densities of the most massive compact system at high-redshift.  
\item Furthermore, the predictions of semi-analytic models that trace the evolution of massive galaxies through minor mergers agree very well with the number densities of the most massive BOSS compacts with the formation redshift of $z>2$. 
\end{itemize}      

A direct measurement (rather than a lower limit) of the number density of massive compact quiescent galaxies at intermediate redshift depends on a redshift survey of a complete photometric sample of compact galaxy candidates. It is important that the survey spans the full range of galaxy environments in the intermediate-redshift universe. Samples of this kind would enable studies of the detailed structure of the compact massive galaxies along with the possible dependence of these properties on the environment. It is very difficult to conduct these investigations at high redshift. Thus intermediate redshift compact galaxy candidates provide an important window on the nature of these systems and their evolution. 

\acknowledgments
We thank the anonymous referee for thoughtful suggestions that improved the clarity of our manuscript. ID is supported by the Harvard College Observatory Menzel Fellowship and the Natural Sciences and Engineering Research Council of Canada Postdoctoral Fellowship (NSERC PDF-421224-2012). The Smithsonian Institution supports the research of HSH, MJG, and IC. IC acknowledges support from grant MD-3288.2012.2. and from the Russian Science Foundation project \#14-22-00041.

We acknowledge the use of the CADC data collections. This work is based on the SDSS-III dataset. Funding for SDSS-III has been provided by the Alfred P. Sloan Foundation, the Participating Institutions, the National Science Foundation, and the U.S. Department of Energy Office of Science. The SDSS-III web site is~\url{http://www.sdss3.org/}. SDSS-III is managed by the Astrophysical Research Consortium for the Participating Institutions of the SDSS-III Collaboration including the University of Arizona, the Brazilian Participation Group, Brookhaven National Laboratory, Carnegie Mellon University, University of Florida, the French Participation Group, the German Participation Group, Harvard University, the Instituto de Astrofisica de Canarias, the Michigan State/Notre Dame/JINA Participation Group, Johns Hopkins University, Lawrence Berkeley National Laboratory, Max Planck Institute for Astrophysics, Max Planck Institute for Extraterrestrial Physics, New Mexico State University, New York University, Ohio State University, Pennsylvania State University, University of Portsmouth, Princeton University, the Spanish Participation Group, University of Tokyo, University of Utah, Vanderbilt University, University of Virginia, University of Washington, and Yale University.


\end{document}